\documentclass [11pt]{article}
\title{Relativistic Rotation: A Comparison of Theories}
\author{by Robert D. Klauber\\1100 University Manor Dr., 38B \\Fairfield, Iowa 52556
\\rklauber[AT]iowatelecom.net or rklauber[AT]netscape.net}

\date{   }

\usepackage{graphicx}
\usepackage{longtable}

\begin{document}

\maketitle
\bigskip
\bigskip
\bigskip

Alternative theories of relativistic rotation considered viable as of 2004 
are compared in the light of experiments reported in 2005. En route, the 
contentious issue of simultaneity choice in rotation is resolved by showing 
that only one simultaneity choice, the one possessing continuous time, gives 
rise, via the general relativistic equation of motion, to the correct 
Newtonian limit Coriolis acceleration. In addition, the widely dispersed 
argument purporting Lorentz contraction in rotation and the concomitant 
curved surface of a rotating disk is analyzed and argued to be lacking for 
more than one reason. It is posited that not by theoretical arguments, but 
only via experiment can we know whether such effect exists in rotation or 
not.

The Coriolis/simultaneity correlation, and the results of the 2005 
experiments, support the Selleri theory as being closest to the truth, 
though it is incomplete in a more general applicability sense, because it 
does not provide a global metric. Two alternatives, a modified Klauber 
approach and a Selleri-Klauber hybrid, are presented which are consistent 
with recent experiment and have a global metric, thereby making them 
applicable to rotation problems of all types.


\bigskip
\bigskip

\section{Introduction}
The recent book \textit{Relativity in Rotating Frames}\cite{Rizzi:2004}, edited by Rizzi and Ruggiero, 
acknowledged the little recognized lack of a universally accepted theory of 
relativistic rotation and presented several different theories by various 
invited contributors. Though what has been termed the traditional theory was 
preferred by most, alternatives were considered viable, given the 
experimental evidence at the time of publication.

Since the Rizzi and Ruggiero text publication, two 
articles\cite{Rizzi:2005}$^{,}$\cite{Selleri:2005} and three 
reports\cite{Antonini:2005}$^{,}$\cite{Stanwix:2005}$^{,}$\cite{Herrmann:2005} 
of relevant experiments have appeared which appear to shed further light on 
this issue. This article reviews differing theories presented in Rizzi and 
Ruggiero in view of these recent developments, and attempts to discern which 
of them remain feasible. In so doing, it also investigates certain 
theoretical matters germane to the subject, and draws related conclusions.

Secs. \ref{sec:mylabel1}, \ref{sec:relevant} and 
\ref{sec:physical} provide background: respectively, the traditional 
approach to relativistic rotation, with emphasis on what I submit to be 
unresolved inconsistencies in that approach; certain relevant experiments 
prior to 2005; and the role of physical components in tensor analysis. 
Though these are background sections, they contain both new perspectives on 
some issues (such as use of the Sagnac experiment in predicting a 
Michelson-Morley result for rotation) and other little known background 
material that is essential to understanding later sections.

Conventionality of synchronization is addressed in Sec. 
\ref{sec:alternative}, with particular attention to the little 
appreciated ambiguity it generates with regard to measures of length. 
Sec. 6 extends the conventionality thesis to 
show that Lorentz contraction, via its standard definition, is different for 
different synchronization choices within the same frame, and thus Lorentz 
contraction can hardly be assumed an absolute, measurable observable 
(particularly for rotation).

Sec. \ref{sec:mylabel2} contains a proof of perhaps the most 
salient conclusion in the article. Coriolis acceleration arises from off 
diagonal time-space components in the rotating frame metric, and those 
components vary directly with choice of simultaneity in the rotating frame. 
Only for the unique choice of simultaneity equal to that of the lab can one 
predict the correct Newtonian limit Coriolis acceleration found in nature. 
This implies conventionality of synchronization is not applicable to 
rotating frames.

Sec. \ref{sec:mylabel3} summarizes the predictions of five 
theories of rotation presented in \textit{Relativity in Rotating Frames} with regard to Michelson-Morley, Lorentz 
contraction, continuity in time, and Coriolis acceleration. Sec. 
\ref{sec:convergence} then notes the underlying convergences between 
ostensibly disparate theories. The requirements for a complete general 
relativistic theory of rotation, in particular, the need for a global 
metric, are discussed in Sec. \ref{sec:mylabel4}.

The results of the three 2005 Michelson-Morley type experiments using 
rotating apparatuses are presented in Sec. \ref{sec:mylabel5}, 
followed by an evaluation in Sec. \ref{sec:evaluation} of the 
proposed theories in light of those results. As none are found completely 
satisfactory, two alternative, suitable analysis approaches are presented in 
Sec. \ref{sec:mylabel6}.

Sec. \ref{sec:loose} ties up loose ends regarding rotating 
disk surface curvature and the Selleri limit paradox. The conclusions are 
summarized in Sec. \ref{sec:summary}.

The article is relatively lengthy, so, as an aid to readers, most sections 
end with a succinct statement of the section's key conclusion(s).

\section{The Traditional Approach}
\label{sec:mylabel1}
\subsection{Method}
\label{subsec:method}
The traditional approach to relativistic rotation assumes one can do what 
has been done successfully in other, non-rotating, but accelerating cases. 
Local co-moving inertial frames (LCIFs) instantaneously at rest relative to 
the accelerating frame are used to measure quantities such as distance and 
time, which would, in principle, be the same as that measured with standard 
meter sticks and clocks by an observer in the rotating frame itself. This is 
often referred to as the ``locality 
principle''\cite{ller:1969}$^{,}$\cite{Stachel:1950}$^{,}$\cite{Mashoon:2003}. 
The local values are typically integrated to determine global values.

An oft cited example, first delineated by Einstein\cite{Ref:1}, is the 
purported Lorentz contraction of the rim of a rotating disk. (Or 
alternatively, the circumferential stresses induced in the disk when the rim 
tries to contract but is restricted from doing so via elastic forces in the 
disk material.) An LCIF instantaneously co-moving with a point on the rim, 
it is argued, exhibits Lorentz contraction of its meter sticks in the 
direction of the rim tangent as observed from the non-rotating frame, via 
its velocity, \textit{v = $\omega $r}. From the locality principle, one then finds the number of 
meter sticks laid end to end around the circumference of the rotating disk 
is increased as observed from the inertial frame, because of the Lorentz 
contraction of the meter sticks. There is no change in radial distances. As 
in SRT, these same meter sticks do not appear contracted to the observer on 
the rotating frame, but their number is the same as that observed from the 
inertial frame. Therefore the rotating frame observer obtains a 
circumference greater than $2 \pi r$, that is, the surface is Riemann 
curved\cite{Relativistic:1975}$^{,}$\cite{Rotating:1977}$^{,}$\cite{Space:1}\footnote{ 
Transforming from the 4D Riemann flat space of the lab to the 4D space of 
the rotating frame implies the 4D space of the rotating frame is also 
Riemann flat. The traditional analysis does not contradict this, but claims 
the 2D surface of the rotating disk is curved within the flat 4D rotating 
frame space.}.

\subsection{Internal Consistency Questions}
\label{subsec:internal}
A number of questions have been raised about the internal consistency of the 
traditional approach. We review some of them here.

The reader should note that the arguments I present herein have been debated 
with both advocates and other questioners of the traditional approach, 
without universal agreement. See Rizzi and Ruggiero [1] Dialogues I, II, 
III, IV, and VI, and many articles therein, as well as 
Weber\cite{Weber:1999}. For a thorough appreciation of the issues, and the 
points and counterpoints made, the reader should review these references.

\subsubsection{The time gap}
\label{subsubsec:mylabel1}
\paragraph{Summary of the Issue\\}
Fig. 1 depicts one of the infinitesimal LCIFs of the traditional approach 
in a global 4D frame (with one spatial dimension suppressed in the global 
frame and two in the LCIF.) Consider the non-rotating (lab) frame as K; the 
rotating (disk) frame as k; and the LCIF as K$_{LCIF}$.

\begin{figure}
\centerline{\includegraphics{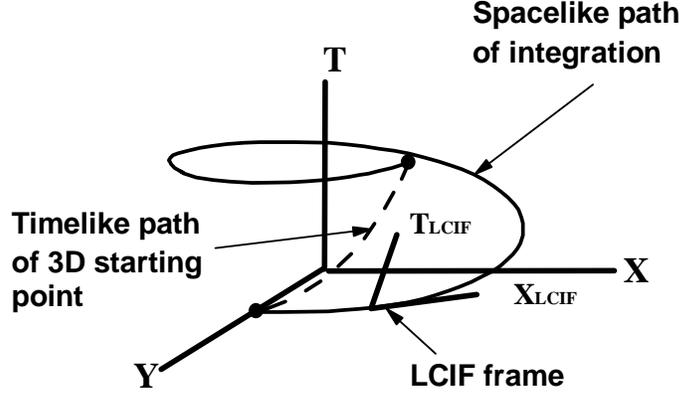}}
\caption{. The Traditional Theory Time Gap}
\label{fig1}
\end{figure}

Note that, via special relativity theory (SRT) with Einstein synchronization 
(we will shortly consider alternative synchronization schemes), events along 
X$_{LCIF}$ which are simultaneous as seen in the LCIF, are not simultaneous 
as seen in K. From the LCIF perspective, each point on the spacelike path 
shown (the integration path for circumferential length determination) is 
simultaneous with every other point on the path. From the lab frame K 
perspective, however, each successive such point is later in time. Thus, if 
one traverses the spacelike path shown, no time should pass in the rotating 
frame between the beginning and ending events, which are both located at the 
same 3D point fixed on the rotating disk rim. Thus, a clock located at this 
3D point on the rim should show the same time for both events. However, from 
the K frame the two events are obviously separated by a timelike interval, 
so the clock cannot show the same time at both events.

This point has been made (in greater detail) by many authors (see 
Weber\cite{Weber:1997} and Klauber\cite{Klauber:2004} for two). As Weber 
notes, the clock on the disk is thus out of synchronization with itself -- a 
rather bizarre situation. Equivalently, since time ``jumps'' a gap between 
360$^{o}$ and 0$^{o}$, one could say time is discontinuous on the rotating 
disk. Also equivalently, one could say time is multivalued, as a given event 
has more than one time associated with it.

\noindent
\paragraph{Traditional Theory Approaches to Resolving the ``Time Gap''\\}
\label{para:traditional}
In recent years, this ``time gap'' problem has been addressed by approaches 
labeled 
``desynchronization''\cite{Rizzi:1999}$^{,}$\cite{Rizzi:2002}$^{,}$\cite{Rizzi:2006} 
and ``discontinuity in 
synchronization''\cite{Cranor:2000}$^{,}$\cite{Anandan:1981}, which 
effectively resynchronize clocks during the course of an experiment, 
depending on the path taken by light during the experiment. Ghosal et 
al\cite{Ghosal:2004} provide one critique of the desynchronization 
approach, and we discuss it in Sec. \ref{subsec:rizzi}.

An older argument purporting to explain the time gap likens it to traveling 
at constant radius in a polar coordinate system, for which the $\phi $ value 
is discontinuous at 360$^{o}$ (or multivalued.) Similar logic has been 
applied for time, with the International Date Line for the time zone 
settings on the earth. If one starts at that line and proceeds 360$^{o}$ 
around the earth, one returns to find one must jump a day in order to 
re-establish one's clock/calendar correctly.

\noindent
\paragraph{Arguments for Physical Interpretation of the ``Time Gap''\\}
\label{para:arguments}
In the desynchronization approach with Einstein synchronization, a given 
event in a particular rotating frame can have any number of possible times 
on it. If one starts at 0$^{o}$ and Einstein synchronizes the clock at 
360$^{o}$ using LCIFs in the ccw direction, one gets a particular time 
setting. If one proceeds in the cw direction, one gets a different setting 
for the same clock. If one travels inward 1 meter, then 360$^{o}$ around, 
then back outward 1 meter, one gets yet another setting for that clock. 
Thus, one has innumerable possible times to choose from for any given event, 
and innumerable different desynchronizations to be made.

This does not happen in translation, where according to traditional SRT, no 
matter what path (straight or curved) one takes to Einstein synchronize 
distant clocks, one always obtains the same time for the same event. There 
are no time discontinuities, and time is not multivalued.

The plethora of possible settings for the same clock in a rotating frame 
results from insisting on ``desynchronization'' of clocks in order to keep 
the (one way) speed of light locally $c$ 
everywhere\cite{Rizzi:2002}$^{,}$\cite{Rizzi:2006}. And thus, one 
is in the position of choosing whichever value for time one needs in a given 
experiment in order to get the answer one insists one must have (i.e., 
invariant, isotropic local light speed.) One can only then ask if this is 
really physics or not. Can an infinite number of possible readings on a 
single clock at a single event for a single method of synchronization be 
anything other than meaningless?

The polar coordinate analogy, the author believes, confuses physical 
discontinuity with coordinate discontinuity. In 2D, place a green X at 
0$^{o}$, travel 360$^{o}$ at constant radius, then place a red X. The red 
and green marks coincide in space. There is no discontinuity in space 
between them, although there is a discontinuity in the coordinate $\phi $.

Flash a green light on the equator at the International Dateline, then trace 
a path once around the equator along which no time passes. If one flashes a 
red light at the end of that path, the red and green flashes are coincident 
in space and time. There is no physical discontinuity, even though time zone 
clocks show a coordinate discontinuity.

In Fig. 1, flash a green light at the beginning of the spacelike path. 
Travel 360$^{o}$ on the disk along the spacelike path (along which no time 
passes according to the traditional analysis), then flash a red light at the 
end. The two lights are not coincident. There is real world invariant 
spacetime gap between them, and they exist at different points in 4D.

\paragraph{Conclusion:} The time discontinuity is physical, not merely coordinate.
Peres was aware of this time discontinuity, calling it a \textit{``heavy price which we are paying to make the [circumferential] velocity of light ... equal to c''}\cite{Peres:1}. 
Dieks\cite{Dieks:1991} noted that though arbitrary in certain senses, time 
in relativity must be \textit{``directly linked to undoubtedly real physical processes''}. I agree.

\noindent 
\paragraph{Alternative Synchronization Conventions\\}
The gauge synchronization philosophy\cite{Anderson:1998} champions 
innumerable, equally valid, synchronization schemes, and some might cite it 
as justification for the multivalued time result discussed above. Yet, 
within any one synchronization scheme, time is single valued and continuous, 
and clocks are all in synchronization with themselves. The resynchronization 
shift is defined along a straight line, and every possible synchronization 
path (using light rays or local LCIFs along curved or straight routes) 
results in the same unique setting for a given clock. For a given 
synchronization scheme, each event within a given frame has a single time 
associated with it, and there are no ``time gaps''. This is not true in the 
desynchronization approach to rotating frames.

Formally, alternative synchronizations can be expressed mathematically as
\begin{equation}
\label{eq1}
dt_{alt} =C(dT_{orig} -\kappa dX_{orig} ),
\end{equation}
where the subscript ``orig'' implies the original, Einstein synchronization 
coordinates; and ``alt'', the new alternative synchronization coordinates. 
Note that $C$ changes the clock rate, but in inertial coordinate systems where 
standard clocks are used, and there is no relative velocity between the 
``orig'' and ``alt'' systems, $C$ = 1. Different values for \textit{$\kappa $} result in 
different alternative simultaneity schemes (i.e., different settings on 
clocks at different $X_{orig}$ locations.) For \textit{$\kappa $} = 0, the original and 
alternate coordinate systems have the same simultaneity (even if $C$ \textit{$\ne $} 1.)

In conventionality of synchronization theory, $X_{orig}$ is typically an axis 
(straight line). When (\ref{eq1}) is assumed to apply to rotation, $X_{orig}$ 
represents the (non-straight line) spacelike path shown in Fig. 1.

In Fig. 1, non-Einstein synchronization gauges can be visualized as having 
different slopes for the spacelike path than that shown. Only when the 
rotating frame shares the same simultaneity as the lab frame does this path 
have zero slope ($\kappa $ = 0, if the lab frame is ``orig''), and only then 
is there no time discontinuity (singled valued time.)

The fact that, on the rotating disk, our synchronization shift direction 
$X_{orig}$ is not a straight line, but ``doubles back'' to its starting 
point, is the proverbial ``fly in the ointment''. Note that were we to apply 
the re-synchronization approach of (\ref{eq1}) within a single inertial reference 
frame, where our direction of synchronization shift was around a 
circumference rather than a straight line, we would find the same problem. 
Our starting and ending clocks would have different settings for the same 
event.

\noindent
\underline {\textbf{Conclusion:}} \textit{The theory of conventionality of synchronization is only self consistent (i.e., has single valued, continuous time) when the re-synchronization shift is defined along a straight line direction.} And hence, it appears inconsistent, and 
thus inappropriate, for use around the rim of a rotating disk.

\subsubsection{Selleri limit discontinuity}
\label{subsubsec:selleri}
Selleri\cite{Selleri:2004}$^{,}$\cite{Selleri:1997}$^{,}$ has used the 
Sagnac\cite{Sagnac:1914}$^{,}$\cite{Post:1967} result to suggest another 
type of discontinuity. Given that global light speeds in the cw and ccw 
directions, as measured by the same clock on the rim of a rotating disk, are 
not equal, then geometric symmetry implies they are locally unequal as well.

The light speed anisotropy is a function of $v=\omega r$. Thus, in the limit 
where radial distance $r \to \infty $, angular rotation $\omega \to $ 
0, such that $\omega r \to $ constant and $\omega ^{2}r \to $ 0, the 
anisotropy would remain non-zero. However, in such a limit, with 
acceleration effectively zero, one would expect the local frame to approach 
an inertial frame, which for Einstein synchronization, would have no such 
anisotropy. Thus, it was argued, there is a discontinuity in light speed, 
whereby anisotropic light speed jumps discontinuously to isotropic light 
speed in the limit.

This issue has an extensive history, as can be gleaned by the references 
cited. The latest volleys were fired by Rizzi and Serafini\cite{Ref:2} and 
Ghosal et al\cite{Ghosal:2004} who noted that the discontinuity 
disappears if one adopts the same synchronization scheme in the limit frame 
as that adopted on the rotating disk rim. Selleri\cite{Selleri:2006} 
generally agreed, but voiced some opposition on philosophic grounds.

However, if, as is championed herein, the only consistent 
synchronization/simultaneity on the rotating disk (including at the limit 
location) is that for which no time discontinuity exists, then the problem 
remains, albeit in slightly different form (see conclusion following.)

\medskip
\noindent
\underline {\textbf{Conclusion:}} If the rotating frame at the limit 
location, by virtue of the locality principle, is effectively equivalent to 
an LCIF, then the rotating frame should be synchronizable in innumerable 
valid ways. But insistence on temporal continuity, and thus unique 
synchronization/simultaneity, restricts it from being so.

I suggest that the assumption of local equivalence of LCIFs in rotating 
frames is not proper, for reasons delineated in Sec. 
\ref{subsec:resolution}

\subsubsection{Lorentz contraction issues}
\label{subsubsec:lorentz}
The traditional analysis (Sec. \ref{subsec:method}) implies that 
the length contraction effect along a rotating disk rim is absolute, in the 
sense that both the lab and disk observer agree that the disk meter sticks 
measure a circumference $C >$ 2$\pi r$. However, it could be argued that according 
to SRT, an observer does not see his own lengths contracting. Only a second 
observer moving relative to the first observer sees the first observer's 
length dimension contracted. Hence, from the point of view of the disk 
observer, her own meter sticks should not be contracted\footnote{ Some may 
contend that other meter sticks in the rotating frame (such as a meter stick 
on the far side of the rim) appear to have velocity with respect to a given 
rim observer. However, an observer fixed on the rim of the disk does \textit{not} see 
the far side of the disk moving relative to her, since she, unlike the local 
Lorentz frame with the same instantaneous linear velocity, is rotating. 
There are two different local frames of interest here. Both have the same 
instantaneous rectilinear velocity as a point on the rim. But one rotates 
relative to the distant stars (at the same rate as the disk itself), and one 
does not. The latter is equivalent to a local Lorentz frame. The former is 
not, and represents the true state of an observer anchored to the disk. The 
difference in kinematics between these two frames is significant, and is 
dealt with in ref. \cite{Klauber:1}.}\cite{Klauber:1}, and there 
should be no curvature of the rotating disk surface.

Tartagli\cite{Tartaglia:1999} and I have made this point independently, and 
Strauss\cite{Strauss:1974} made a closely related one. In response to 
Strauss, Gr{\o}n\cite{Rotating:1977} noted that in accelerating the disk 
from a non-rotating state, the required local accelerations of rim points 
would force the distance between those points to be non-Born rigid, and this 
may appear to be a reasonable resolution of the issue. However, a subsequent 
refinement of the argument by Gr{\o}n, positing the effective 
``gravitational'' field on the disk responsible for the $C >$ 2$\pi r$ value found 
by a disk fixed observer, appears problematic to me. Interested readers 
should see Ref \cite{Rizzi:2004}, Dialogue VI, pp 444-445.

Yet, the central point made by Tartaglia and me is that in SRT, with a 
relative velocity difference between two observers, neither would see any 
difference at all in the measurements each would make in his own frame from 
when they were stationary with respect to one another. So our key question 
is why should a disk observer measure anything different at all with her 
meter sticks when she is moving with respect to the lab, from what she 
measures when is she is not so moving?

In a related argument, consider the disk observer looking out at the meter 
sticks at rest in the lab and seeing them as having a velocity with respect 
to her. Hence, by the traditional logic, she sees them as contracted and 
must therefore conclude that the lab surface is curved, which of course, it 
is not. The traditional analysis thus appears inconsistent.

Gr{\o}n also responds to this in Ref \cite{Rizzi:2004} [Dialogue III, 
pp. 411-421], though from my perspective, found therein, one can glean that 
I am less than convinced.

A corollary to the foregoing conundrum exists. As the earth rotates, stars 
distant from us travel, relative to us on the earth surface, at speeds close 
to (and even exceeding!) the speed of light as seen in our frame. Yet there 
is no Lorentz contraction of those stars, even though the locality principle 
suggests we should. Thus, simple-minded application of standard SRT Lorentz 
contraction directly to rotation is obviously wrong.

\medskip
\noindent
\underline {\textbf{Conclusion:}} If Lorentz contraction exists in a 
rotating frame, it is argued that it does not arise via the same mechanism 
as in traditional SRT. (See also, Sec. 6.)

\section{Relevant Experiments}
\label{sec:relevant}
\subsection{Sagnac: A Clue to Michelson-Morley}
\label{subsec:sagnac}
The celebrated 
Sagnac\cite{Sagnac:1914}$^{,}$\cite{Post:1967}$^{,}$\cite{Klauber:2003} 
experiment has, no doubt, generated more discussion than any other aspect of 
relativistic rotation. We will not repeat the common arguments here, but 
present what I believe to be a novel look at, and a new conclusion with 
regard to, it.

In the Sagnac experiment, light rays are propagated in opposite directions 
around an effectively circumferential path within a rotating frame, with one 
ray returning to the emission point prior to the other. Some\cite{See:1} 
have taken this to imply anisotropic local one-way light speed in rotating 
frames, though others have argued against this. The most popular recent 
position is that it all depends on one's choice of synchronization (or more 
correctly, simultaneity), though as noted above, I seriously question 
whether, for rotation, a multiplicity of such choices exists.

There is another issue, however, which the Sagnac experiment can help us 
understand -- the Michelson-Morley (MM) experiment in a rotating frame. Due 
to the universal agreement on the exact expression for the time difference 
between arrival times of clock-wise (cw) and counterclockwise (ccw) 
circularly propagating light rays in a rotating frame, we can determine the 
time for back and forth light propagation locally in the circumferential 
direction as measured in the rotating frame.

\medskip
\noindent
\underline {Analyzing Sagnac Time Difference}

Using upper case letters to represent lab and lower case for the rotating 
frame, the time taken for the cw light pulse to traverse a circumference in 
the rotating frame of radius $R$ as measured from the lab is $T_{1}$; the time 
for the ccw pulse is $T_{2}$. Then from the lab frame, the lengths traveled 
in each case are
\begin{equation}
\label{eq2}
cT_1 =2\pi R-\omega RT_1 \,\,\,\,\mbox{cw}\, ,\,\,\,\,\,\,\,cT_2 =2\pi 
R+\omega RT_2 \,\,\,\,\,\,\mbox{ccw},
\end{equation}
where $\omega $ is the angular velocity as seen from the lab.

Solving (\ref{eq2}) for $T_{1}$ and $T_{2}$, we find the Sagnac experiment time 
difference as measured in the lab to be
\begin{equation}
\label{eq3}
T_{Sagnac} =T_2 -T_1 =\frac{\left( {2\pi R} \right)\left( {2\omega R} 
\right)}{c^2\left( {1-v^2/c^2} \right)}.
\end{equation}
where \textit{v=$\omega $R}. Since we know the time in the rotating frame is slowed by the 
Lorentz factor, this time, measured by a physical clock fixed in the 
rotating frame, is
\begin{equation}
\label{eq4}
t_{Sagnac,phys} =T_{Sagnac} \sqrt {1-v^2/c^2} =\frac{\left( {2\pi R} 
\right)\left( {2\omega R} \right)}{c^2\sqrt {1-v^2/c^2} }.
\end{equation}
The above analysis is well known. We have presented it only as background 
for the following.

\medskip
\noindent
\underline {Analyzing Michelson-Morley Result in Rotating Frame}

Now, instead of considering two light flashes released at the same time in 
opposite directions, consider one cw flash that travels 360$^{o}$ in the 
rotating frame, then is reflected and returns in a ccw direction to its 
point of origin. The same relations (\ref{eq2}) apply, so that the round trip time 
measured in the lab is
\begin{equation}
\label{eq5}
T_{RT,circum} =T_1 +T_2 =\frac{2(2\pi R)}{c\left( {1-v^2/c^2} \right)}.
\end{equation}
And the round trip time measured by a physical clock at the 
emission/reception point in the rotating frame is
\begin{equation}
\label{eq6}
t_{RT,phys,circum} =T_{RT,circum} \sqrt {1-v^2/c^2} =\frac{2(2\pi R)}{c\sqrt 
{1-v^2/c^2} }.
\end{equation}
Note that if Lorentz contraction exists in the rotating frame, the 
circumferential distance $l$ measured with meter sticks in the rotating frame 
increases such that it is greater than 2$\pi R$ by the Lorentz factor, i.e., 
$l=2\pi R/\sqrt {1-v^2/c^2} $. Then,
\begin{equation}
\label{eq7}
t_{RT,phys,circum} =\frac{2l}{c}\,\,\,\,\,\,\mbox{(if Lorentz contraction 
exists)},
\end{equation}
and the round trip speed of light is $c$. On the other hand, if Lorentz 
contraction does not exist, then $l =$ 2$\pi R$,
\begin{equation}
\label{eq8}
t_{RT,phys,circum} \cong \frac{2l}{c}\left( {1+\frac{1}{2}\frac{v^2}{c^2}} 
\right)\,\,\,\,\,\,\,\,\,\,\,\mbox{(if Lorentz contraction does not exist)}
\end{equation}
and the round trip light speed differs from $c$ at second order by the familiar 
$\beta $ factor.

For distances less than one complete circumferential traversal in the 
rotating frame, the same kinematics exist. For example, for a 1$^{o}$ back 
and forth path, the same relations as (\ref{eq7}) or (\ref{eq8}) apply, but $l$ would then be 
1/360$^{th}$ the distance.

Thus, the theoretically derived, and universally agreed to, values of 
traversal times found via (\ref{eq2}) in the Sagnac analysis can be employed to 
analyze a local MM experiment.

In a MM experiment, one must compare round trip light times in perpendicular 
directions, and thus we must also deduce the radial (or alternatively, the 
$Z$ direction) round trip light time. From the classic elementary 
analysis\cite{Born:1965} of the MM experiment, we know the round trip time 
to and from a point fixed in the rotating frame, in a direction 
perpendicular to the circumference, measured by a lab clock is
\begin{equation}
\label{eq9}
T_{RT,\bot } =\frac{2(\Delta R)}{c\sqrt {1-v^2/c^2} },
\end{equation}
where $\Delta R$ here could actually be any distance in the $R-Z$ plane between 
the emission (reception) and reflection points. Given the known time 
dilation in the rotating frame, and the agreed upon lack of Lorentz 
contraction in a direction perpendicular to velocity, we have
\begin{equation}
\label{eq10}
t_{RT,phys,\bot } =T_{RT,\bot } \sqrt {1-v^2/c^2} =\frac{2(\Delta 
R)}{c}=\frac{2l}{c}.
\end{equation}
Thus, the round trip speed of light in any direction orthogonal to the 
rotating frame local velocity is $c$. (A more sophisticated analysis leading to 
this result can be found in Ref. \cite{Klauber:2005}, p. 130.) 

\noindent
\underline {\textbf{Conclusions}}\textbf{: }

\begin{enumerate}
\item A local, rotating-frame-fixed MM experiment of sufficient accuracy, comparing light travel times in perpendicular directions, can determine which of (\ref{eq7}) or (\ref{eq8}) is valid, i.e., whether Lorentz contraction exists in a rotating frame or not.
\item The highest local velocity (and thus the most sensitive experiment) due to rotation we could practically obtain would be from using the earth itself as our rotating frame, for which the local velocity is of the order .35 km/sec.
\end{enumerate}
\subsection{Brillet and Hall}
\label{subsec:brillet}
In 1978, in a milestone experiment, Brillet and Hall\cite{Brillet:1979} 
used a Fabry-Perot interferometer to confirm to high order (10$^{-15}$) the cosmic isotropy of the speed of light.  Because their apparatus rotated with respect to the earth fixed-frame, the data taken could also serve as a MM type test that was, for the first time, sensitive enough to measure two-way anisotropy of light speed that might be due to the earth's rotation.  This earth-fixed frame data displayed a
variation between perpendicular round trip 
travel times of 2X10$^{-13}$, which is very close to the 
$\raise.5ex\hbox{$\scriptstyle 1$}\kern-.1em/ 
\kern-.15em\lower.25ex\hbox{$\scriptstyle 2$} v^{2}$/$c^{2}$ of (\ref{eq8}), with 
$v$ = earth surface speed about its axis at the test location. This signal had 
effectively constant phase in the earth-fixed frame, and was ``persistent''.
(See Sec. 11 for an explanation of how this earth frame signal effectively 
averaged out to zero for the cosmic light speed isotropy result.)   

Most subsequent researchers have considered this signal to be anomalous, 
though 
others\cite{Bel:2004}$^{,}$\cite{Klauber:2004}$^{,}$\cite{Klauber:2005} 
have suggested that something significant may be behind it, i.e., true light 
speed anisotropy due to idiosyncrasies of rotating frames. 

\subsection{Other Experiments}
\label{subsec:other}
A large number of tests of relativity theory have been carried out since the 
Michelson-Morley experiments in the 1880s. All relevant ones found by this 
author prior to the time (2004) of publication of ref. 
\cite{Klauber:2004} are listed therein in Sec. 3, where it is noted 
that before 2005, none other than that by Brillet-Hall have had both 1) high 
enough sensitivity and 2) a suitable design (such as rotating apparatus 
relative to the earth fixed frame) to permit detection of a non-null MM type 
signal within a rotating frame.

After discussing certain requisite theoretical points, and summarizing 
various theories of relativistic rotation that have been proposed, we will 
review results from three experiments completed in 2005 that have shed some 
light on this question.

\medskip
\noindent
\underline {\textbf{Conclusion:}} Prior to 2005, there was no experimental 
evidence and no iron-clad theoretic reason to conclude that the anomalous 
Brillet and Hill earth-fixed frame signal was not indicative of a true round trip light speed 
anisotropy in rotating frames.

\begin{figure}
\centerline{\includegraphics{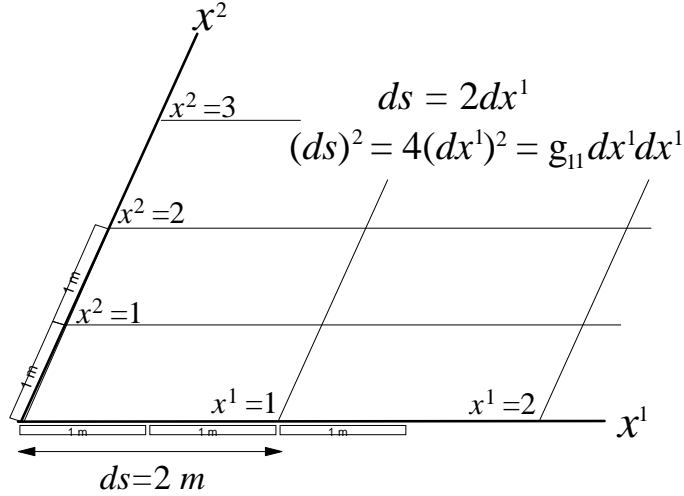}}
\caption{. Physical Length vs Coordinate Value}
\label{fig2}
\end{figure}

\section{Physical Components and Generalized Tensor Analysis}
\label{sec:physical}
\subsection{Differential Geometry Definitions}
To facilitate clarification of salient points to be discussed below, we 
briefly review definitions of the interval, coordinate components, and 
physical components, according to differential geometry (alternatively 
called generalized tensor analysis). The interval \textit{ds}, as shown via the two 
dimensional spatial space depicted in Fig. 2, can be considered equal to 
the number of meter sticks between the endpoints of \textit{ds}. It is a very physical, 
measured quantity. Coordinates like \textit{dx}$^{1}$, on the other hand, are mere 
numbers, reflecting labels attached to arbitrarily laid out coordinate mesh 
lines. The metric, of course, relates coordinate values (mere labels) to the 
interval (physically measured value).
\begin{equation}
\label{eq11}
\left( {ds} \right)^2=g_{ij} dx^idx^j=\,\,\mbox{(number of physical meter 
sticks)}^2
\end{equation}
And for vectors in general, where we use velocity as an example,
\begin{equation}
\label{eq12}
{\rm {\bf v}}\cdot {\rm {\bf v}}=g_{ij} v^iv^j=\,\,\mbox{(physically 
measured velocity)}^2.
\end{equation}
If we look at one component of a vector, such as \textit{dx}$^{1}$ in Fig. 2, it has 
a coordinate value (\textit{dx}$^{1}$ =1 in the figure) and a physically measured value 
(\textit{ds} along the $x^{1}$ direction = 2 meters.) Note that in Fig. 2, $g_{{\rm 
x}{\rm x}}$= 4, and along the direction shown,
\begin{equation}
\label{eq13}
\left( {ds} \right)^2=g_{11} dx^1dx^1=4(\ref{eq1})(\ref{eq1})=\,\,\mbox{(meter sticks along 
}x^1)^2,
\end{equation}
so two meter sticks cover a coordinate distance of one unit. The physical 
distance of two meter sticks in the $x^{1}$ direction is called the \textit{physical component}\cite{The:1972}, 
designated $d\hat {x}^1$, and in general,
\begin{equation}
\label{eq14}
ds_{\mbox{along}\,\mbox{axis i}} =d\hat {x}^i=\sqrt {g\underline{_{ii} }} 
dx^i=\,\,\mbox{physically measured value},
\end{equation}
where underlining implies no summation. A physical component for 
displacement is a special case of proper distance, where that proper 
distance is measured along a coordinate axis.

Similarly, any general vector has physical components associated with it, 
and thus, we can express it in two ways, via coordinate components 
(mathematical entities) or physical components (entities measured with 
instruments). That is,
\begin{equation}
\label{eq15}
{\rm {\bf v}}=v^1{\rm {\bf e}}_1 +v^2{\rm {\bf e}}_2 =\hat {v}^1{\rm {\bf 
\hat {e}}}_1 +\hat {v}^2{\rm {\bf \hat {e}}}_2 
\end{equation}
where the carets designate physical components and unit vectors. We caution 
that tensor analysis must be done with coordinate components, not physical 
components\footnote{ To be precise, since physical components are special 
case anholonomic components, they can be used in analyses for which proper 
care is taken to employ the appropriate, and more cumbersome, anholonomic 
mathematical machinery. See Ref. 
\cite{Misner:1973}.}\cite{Misner:1973}, though we need physical 
components to relate analytical results to measured results. Relation (\ref{eq14}) 
allows us to pass back and forth between the two representations.

We note further that relation (\ref{eq14}) is valid for any vector in flat or curved 
space (as it is a local relation), applies to orthogonal or non-orthogonal 
coordinate grids, and has a similar sister expression for tensor components.

For some quite regrettable reason, physical components are rarely mentioned 
in general relativity courses or texts. Thus, the need to review them 
herein.

\subsection{Schwarzchild Example}
\label{subsec:schwarzchild}
Consider a displacement vector expressed via Schwarzchild coordinates where 
\textit{dt} = \textit{d$\theta $} = 0, and $c$ = 1. Then,
\begin{equation}
\label{eq16}
ds^2=\frac{dr^2}{1-2M/r}+r^2d\phi ^2,
\end{equation}
and the vector can be expressed in two different ways, 
\begin{equation}
\label{eq17}
d{\rm {\bf x}}=dr{\kern 1pt}{\rm {\bf e}}_r +\,d\phi {\rm {\bf e}}_\phi 
\,=\,\underbrace {\frac{dr}{\sqrt {1-2M/r} }}_{d\hat {r}}{\rm {\bf \hat 
{e}}}_r \,+\underbrace {\,rd\phi }_{d\hat {\phi }}{\kern 1pt}{\rm {\bf \hat 
{e}}}_\phi .
\end{equation}
Thus, the part of the displacement measured using meter sticks in the 
\textbf{r} direction is the physical component $dr/\sqrt {1-2M/r} $, not the 
coordinate component \textit{dr}.

\section{Alternative Synchronizations and Physical Components}
\label{sec:alternative}
With the foregoing as background, one can appreciate an idiosyncratic 
characteristic of non-Einstein synchronization schemes.

\subsection{Purely Spatial Analogy}
Consider a Cartesian $X-Y$ coordinate grid, where coordinate labels indicate 
physical distance in meters from the origin. Transform to the $x-y$ coordinate 
grid via
\begin{equation}
\label{eq18}
\begin{array}{l}

x=X\,\,\,\,\,\,\,\,\,\,\,\,\,\,\,\,\,\,\,\,\,\,\,\,\,\mbox{Where}\,\,x\,\mbox{is}\,\mbox{forced}\,\mbox{to}\,\mbox{equal}\,\mbox{number} 
\\ 
 y=Y-\kappa 
X\,\,\,\,\,\,\,\,\,\,\,\,\,\,\,\,\,\,\,\mbox{of}\,\mbox{meter}\,\mbox{sticks}\,\,\mbox{along}\,x\,\mbox{axis.} 
\\ 
 \end{array}
\end{equation}
In normal differential geometry, the new coordinates $x$ and $y$ would be mere 
labels, generally not equal to physical distance from the origin. However, 
let us impose the artificial restriction that the coordinates \textit{must} indicate the 
number of meter sticks along the new coordinate axes from the origin. This 
(soon to be appreciated, strange) transformation is displayed in Fig. 3.

\begin{figure}
\centerline{\includegraphics{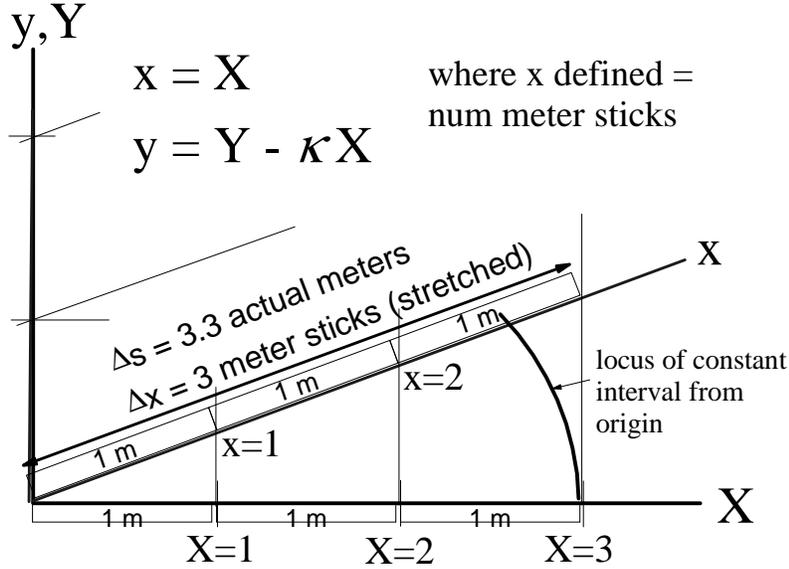}}
\caption{. Artificial (Weird) Spatial Transformation}
\label{fig3}
\end{figure}

Note that to enforce this restriction, the meter sticks along the $x$ axis must 
be stretched, i.e., a meter stick no longer measures one meter of actual 
length. The number of meter sticks for $\Delta s$ shown in the figure is 
three, but the physical length of $\Delta s$ is \textit{not} this, but 3.3 meters. This, 
no doubt, seems bizarre and certainly runs counter to one's normal 
understanding and application of differential geometry. However, as we will 
show below, this is precisely the sort of thing that happens in the theory 
of conventionality of synchronization.

\subsection{Conventionality of Synchronization}
\label{subsec:conventionality}
Consider a resynchronization such as that described by Anderson, 
Vetharaniam, and Stedman\cite{Anderson:1998} (``AVS synchronization''), 
i.e., (\ref{eq1}) with $C$ = 1, analogous to taking the spatial $y,Y$ axes in (\ref{eq18}) over to 
temporal axes \textit{ct} and\textit{ cT}, and for which $X,T$ comprise an Einstein synchronized 
coordinate system. This is
\begin{equation}
\label{eq19}
\begin{array}{l}

x=X\,\,\,\,\,\,\,\,\,\,\,\,\,\,\,\,\,\,\,\,\,\,\,\,\,\,\,\,\,\,\mbox{where}\,\,x\,\,\mbox{is}\,\mbox{constrained}\,\mbox{to}\,\mbox{equal}\,\mbox{the}\, 
\\ 
 ct=cT-\kappa 
X\,\,\,\,\,\,\,\,\,\,\,\,\,\,\,\mbox{number}\,\mbox{of}\,\mbox{meter}\,\mbox{sticks}\,\,\mbox{along}\,x\,\mbox{axis.} 
\\ 
 \end{array}
\end{equation}
Like the prior example with purely spatial coordinates, this transforms 
orthogonal coordinate axes to non-orthogonal coordinate axes (see Fig. 4), 
which in this case are termed non-\textit{time}-orthogonal. From Fig. 4, one can glean 
that a greater value of \textit{$\kappa $} implies a greater slope of the spatial axis $x$ from 
the horizontal.

For later reference, we note that, as shown by AVS, the metric for the 
resynchronized \textit{ct-x} coordinates is
\begin{equation}
\label{eq20}
g_{\alpha \beta } =\left[ {{\begin{array}{*{20}c}
 {-1} \hfill & 0 \hfill & \kappa \hfill & 0 \hfill \\
 0 \hfill & 1 \hfill & 0 \hfill & 0 \hfill \\
 \kappa \hfill & 0 \hfill & {1-\kappa ^2} \hfill & 0 \hfill \\
 0 \hfill & 0 \hfill & 0 \hfill & 1 \hfill \\
\end{array} }} \right],
\end{equation}
where for convenience in comparison with other metrics presented herein, we 
have moved the terms in \textit{$\kappa $} from the second row/column to the third (i.e., we 
have exchanged $x$ and $y$ directions in (\ref{eq20}).)

Note that in such resynchronization transformations, the $x$ coordinate in the 
new coordinate grid is equal to the number of meter sticks from the origin 
along the $x$ axis. However, this is \textit{not} equal to the actual interval length 
$\Delta s$, and it is the latter, not the former, which in differential 
geometry is considered to be the actual, physically measurable 
length\cite{See:2}.

See Fig. 4, noting that, unlike the purely spatial transformation of 
Fig. 3, the presence of a minus sign in the temporal component of the 
metric means the locus of constant interval length is a hyperbola rather 
than a circle. Thus, the actual interval distance along the $x$ axis is \textit{less} than 
the number of meter sticks. Thus, in a seemingly weird way, lengths measured 
with physical meter sticks are \textit{not physical} lengths in the usual differential geometry 
sense.

\begin{figure}
\centerline{\includegraphics{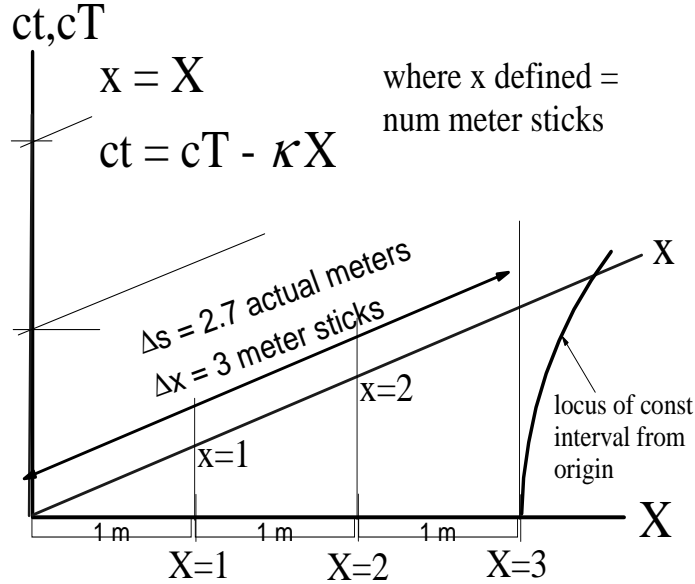}}
\caption{. Synchronization (Space-Time) Transformation}
\label{fig4}
\end{figure}

In spacetime (relativistic) theory, proper length is defined as the interval 
$\Delta s$ along an axis of simultaneity (e.g., the $x$ axis in Fig. 4) and 
this $\Delta s$ corresponds to the term physical length. With Einstein 
synchronization, ``proper length'' is synonymous with ``rest length'' (the 
number of meter sticks laid out between, and at rest with respect to, two 3D 
points.) However, for non-Einstein synchronization, proper length$\Delta 
s_{t=0} =\Delta \hat {x}$ is \textit{not} equal to rest length $\Delta x$.

This is a result of the requirement, in conventionality of synchronization 
theories, that one simply resets clocks at locations separated by constant, 
meter stick measured, distances. Thus, the values assigned to $x$ in such 
theories are no longer mere labels with little physical significance, but 
become pegged to specific values measured with physical instruments in the 
real world. 

The distinction between proper length (actual physical length in 4D 
differential geometry sense) and rest length (physical in terms of using 
meter sticks to measure, which we will henceforth call ``real world'' 
length) will become important when we investigate arguments for, and 
against, curvature of the surface of a rotating disk.

\medskip
\noindent
\underline {\textbf{Conclusion:}} For non-Einstein synchronizations, rest 
length measured via standard real world meter sticks (equal to \textit{dx}) does not 
equal proper length (equal to \textit{ds} along the $x$ axis) in meters. Care must be 
taken in interpreting differential geometry in 4D, as this \textit{ds} is defined 
therein as the ``physical component'' in the $x$ direction, which in a purely 
spatial space equals the number of meter sticks needed to traverse it. In 
4D, this is only true for Einstein synchronization.

\section{Lorentz Contraction and Conventionality of Synchronization}
\label{sec:lorentz}
\begin{figure}
\centerline{\includegraphics{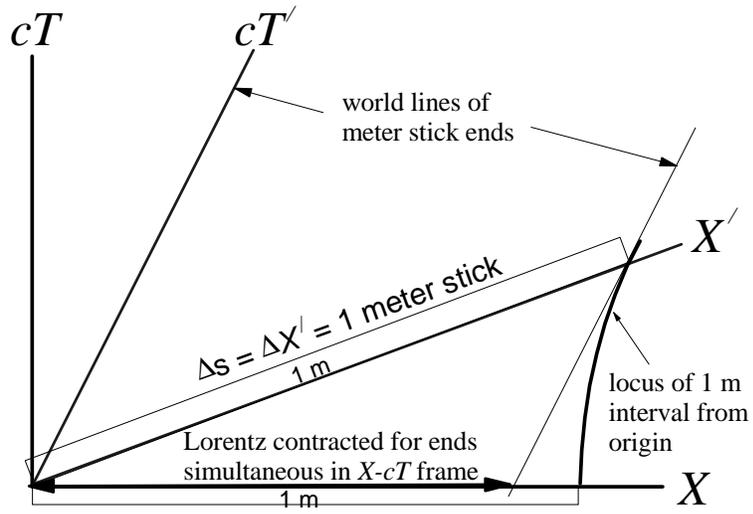}}
\caption{. Lorentz Contraction with Einstein Synch}
\label{fig5}
\end{figure}

Fig. 5 serves as a review of the source of Lorentz contraction between two 
inertial frames in relative motion. The meter stick fixed in the primed 
frame (with the world lines of its endpoints shown) appears to be moving to 
an observer in the unprimed frame. Lorentz contraction arises by stipulating 
that the endpoints of the primed meter stick must be considered simultaneous 
to the unprimed observer. Thus, with this stipulation, and as illustrated, 
the unprimed frame observer determines that the moving meter stick is 
shorter than his own stationary (relative to him) meter stick. The 
intersection of the moving meter stick endpoint worldlines with the 
stationary frame spatial axis determines the amount of the contraction. 
Fig. 5 illustrates this for Einstein synchronized coordinates.

However, as noted in Sec. \ref{subsec:conventionality}, the slope of the 
(unprimed) spatial axis depends on the observer's choice of simultaneity. 
But this means the intersection point of the moving meter stick's right end 
worldline with the unprimed spatial axis will be different for different 
simultaneities in the unprimed frame (see Fig. 6), and thus the length 
seen in the unprimed frame would depend on the simultaneity chosen therein. 

\begin{figure}
\centerline{\includegraphics{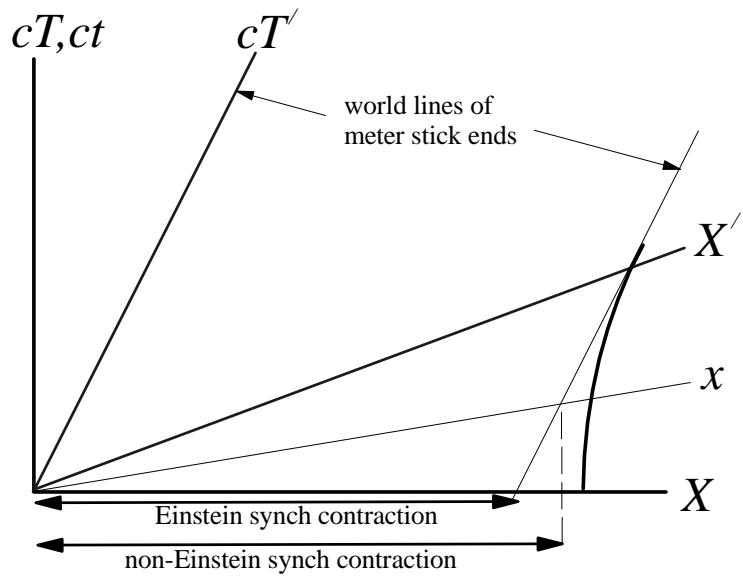}}
\caption{. Non-Einstein Synch Contraction}
\label{fig6}
\end{figure}

Thus, if simultaneity is truly conventional, then so is Lorentz contraction. 
And it follows that, if one insists measurement can be made via Lorentz 
contracted meter sticks on a rotating disk, there can then be no unique 
value for the circumference of that disk, nor any unique curvature for the 
disk surface, as claimed in the traditional approach to relativistic 
rotation.

\medskip
\noindent
\underline {\textbf{Conclusions:}} 

\begin{enumerate}
\item Different synchronization (simultaneity) means different Lorentz contracted lengths for the same moving meter stick (using the standard definition of Lorentz contraction.)
\item Conventionality of synchronization and the traditional rotating disk argument for Lorentz contraction leading to a unique, invariant disk curvature cannot both be true.
\end{enumerate}
\section{A Complete Analysis: Acceleration Included}
\label{sec:mylabel2}
A complete general relativistic analysis of a particular problem is not 
limited to times, lengths, and velocities, but includes accelerations, 
forces, and curvature as well. Central to all of this is the metric, which 
one employs in tensor analysis to determine all of these quantities.

\subsection{Well Known Examples}
\subsubsection{Schwarzchild field}
Given a suitable metric for a Schwarzchild field (abbreviated form shown in 
(\ref{eq16})), one can determine not only physically measured (proper) distances 
between points (see (\ref{eq17}), for example), proper times, and physical 
velocities, but (physical) accelerations of particles, forces, 
electromagnetic relations, and the Riemann tensor, as well.

\subsubsection{Rectilinear acceleration}
As shown in Chapter 6 of Misner, Thorne, and Wheeler\cite{Misner:1973}, 
LCIFs can be used to derive a suitable metric for a frame accelerating in a 
straight line. Using that metric, one can find all the aforementioned 
relevant quantities, kinematic, dynamic, and geometric for that frame.

\subsubsection{Rotation via transformation theory}
\label{subsubsec:rotation}
In rotation, any viable analytic approach must be able to predict 
centrifugal and Coriolis accelerations, as well as curvature. Using the most 
widely employed transformation between the lab and a rotating frame, these 
things can indeed be determined. This analysis procedure is known, but 
relatively difficult to find in the literature, and as it is relevant to 
comments to be made later, I include it here for reference.

For rotation, with familiar symbols for cylindrical coordinates and the 
coordinate transformation
\begin{equation}
\label{eq21}
\begin{array}{l}
 ct=cT\quad \quad \quad \quad \quad \mbox{(a)} \\ 
 r=R\quad \quad \quad \quad \;\;\;\quad \mbox{(b)} \\ 
 \phi =\Phi -\omega T\quad \quad \quad \;\,\mbox{(c)} \\ 
 z=Z\quad \quad \quad \quad \quad \;\;\;\mbox{(d)} \\ 
 \end{array},
\end{equation}
where upper case refers to the lab frame, and lower case to the rotating 
frame, the metric and its inverse\cite{Klauber:2}\footnote{ This metric, 
and all to follow in subsequent sections, can be checked by using it to 
write out the line element for the rotating frame, substituting the 
differential form (in terms of \textit{dt, dr,} etc) of the transformation ((\ref{eq21}) in this 
case), and noting that the resultant is the correct expression of the line 
element in the non-rotating frame.} are
\begin{equation}
\label{eq22}
g_{\alpha \beta } =\left[ {{\begin{array}{*{20}c}
 {-(1-\textstyle{{r^2\omega ^2} \over {c^2}})} \hfill & 0 \hfill & 
{\textstyle{{r^2\omega } \over c}} \hfill & 0 \hfill \\
 0 \hfill & 1 \hfill & 0 \hfill & 0 \hfill \\
 {\textstyle{{r^2\omega } \over c}} \hfill & 0 \hfill & {r^2} \hfill & 0 
\hfill \\
 0 \hfill & 0 \hfill & 0 \hfill & 1 \hfill \\
\end{array} }} \right]\quad g^{\alpha \beta }=\left[ {{\begin{array}{*{20}c}
 {-1} \hfill & 0 \hfill & {\textstyle{\omega \over c}} \hfill & 0 \hfill \\
 0 \hfill & 1 \hfill & 0 \hfill & 0 \hfill \\
 {\textstyle{\omega \over c}} \hfill & 0 \hfill & {(1-\textstyle{{r^2\omega 
^2} \over {c^2}})/r^2} \hfill & 0 \hfill \\
 0 \hfill & 0 \hfill & 0 \hfill & 1 \hfill \\
\end{array} }} \right].
\end{equation}
Note from the form of (\ref{eq21})(a), that for this transformation, the lab and the 
rotating frame have the same simultaneity (\textit{$\kappa $} = 0 in (\ref{eq1}), so that if $\Delta 
T$ = 0 between any two events, then $\Delta t$ = 0). Also, $t$ turns out to be a 
coordinate clock time in the rotating frame, not standard (physical, real 
world) clock time (which varies from $t$ by the Lorentz factor.)

For (\ref{eq22}), the only non-zero Christoffel symbols, found from
\begin{equation}
\label{eq23}
\Gamma _{\alpha \beta \gamma } =\textstyle{1 \over 2}\left( {g_{\alpha \beta 
,\gamma } +g_{\alpha \gamma ,\beta } -g_{\beta \gamma ,\alpha } } 
\right),\,\,\,\,\,\,\Gamma ^\alpha _{\beta \gamma } =g^{\alpha \mu }\Gamma 
_{\mu \beta \gamma } \,,
\end{equation}
are
\begin{equation}
\label{eq24}
\Gamma ^r_{tt} =\Gamma ^1_{00} =-\frac{\omega 
^2r}{c^2}, \,\,\,\,\,\,\,\,\Gamma ^\phi _{tr} =\Gamma ^2_{01} =\frac{\omega 
}{cr}, \,\,\,\,\,\,\,\,\,\Gamma ^r_{t\phi } =\Gamma ^1_{02} =-\frac{\omega 
r}{c}.
\end{equation}
The equation of motion for a geodesic particle, in rotating frame 
coordinates, is
\begin{equation}
\label{eq25}
\frac{d^2x^\alpha }{d\tau ^2}+\Gamma ^\alpha _{\beta \gamma } \frac{dx^\beta 
}{d\tau }\frac{dx^\gamma }{d\tau }=0.
\end{equation}
The relevant 4-velocities are
\begin{equation}
\label{eq26}
\begin{array}{l}
 u^0=\frac{dx^0}{d\tau }=\frac{cdt}{d\tau }=\frac{c}{\sqrt {1-({v}'/c)^2} } 
\\ 
 u^1=\frac{dx^1}{d\tau }=\frac{dr}{d\tau } \\ 
 u^2=\frac{dx^2}{d\tau }=\frac{d\phi }{d\tau }\,\,, \\ 
 \end{array}
\end{equation}
where ${v}'$ is the velocity of the particle as seen in the lab frame (since 
\textit{dt = dT} in the first line of (\ref{eq26}).)

\medskip
\noindent
\underline {Radial Direction Acceleration}

For the $x^{1}=r$ direction, the equation of motion (\ref{eq25}) becomes
\begin{equation}
\label{eq27}
a^r=a^1=\frac{\omega ^2r}{\left( {1-({v}'/c)^2} \right)}+\frac{\omega 
r}{\sqrt {1-({v}'/c)^2} }\frac{d\phi }{d\tau }=\underbrace {\frac{\omega 
^2r}{\left( {1-({v}'/c)^2} \right)}}_{Centrifugal\,\,accel}+\underbrace 
{\frac{\omega \hat {u}^\phi }{\sqrt {1-({v}'/c)^2} }}_{Coriolis\,\,accel}
\end{equation}
where $\hat {u}^\phi =\sqrt {g_{\phi \phi } } d\phi /d\tau =rd\phi /d\tau $ 
is the physical (i.e., measured in m/s using standard meter sticks) 
four-velocity of the particle in the \textit{$\phi $} direction relative to the rotating 
frame. Since the particle is undergoing geodesic motion, as seen from the 
rotating frame, there is acceleration relative to the rotating frame 
coordinates. For a particle fixed at constant radius $r$ in the rotating frame, 
centrifugal and Coriolis pseudo forces equal to the mass times the terms on 
the RH side of (\ref{eq27}) would appear to arise.

The reader who has worked out (\ref{eq24}) to (\ref{eq27}) realizes that the centrifugal 
acceleration arises from the $g_{tt}$ term in the metric, whereas the zeroth 
order Coriolis term in (\ref{eq27}) is due solely to the first order time-space 
component g$_{t\phi }$ in the metric (\ref{eq22}). Without that off-diagonal 
component due to non-time-orthogonal (NTO) rotating coordinates, we would 
find no measurable Coriolis acceleration in rotation.

Further, had the simultaneity chosen been other than that of (\ref{eq21})(a) [i.e., 
with $t$ dependent not only on $T$, but on \textit{$\Phi $} as well], then we would have had a 
different g$_{t\phi }$, and thus a different Coriolis acceleration at zeroth 
order. The Coriolis Newtonian acceleration found in a myriad of experiments 
and applications is the zeroth order approximation of the value shown in 
(\ref{eq27}). Thus, I submit this to be clear evidence that Nature herself has 
decided there is no conventionality of simultaneity/synchronization in 
rotation. The only simultaneity choice producing the known Coriolis effect 
is that of the lab.

\medskip
\noindent
\underline {Tangential Direction Acceleration}

For the $x^{2}$ = \textit{$\phi $} direction, the equation of motion (\ref{eq25}) becomes
\begin{equation}
\label{eq28}
a^\phi =a^2=\frac{d^2\phi }{d\tau ^2}=-\frac{\omega }{r}\frac{dt}{d\tau 
}\frac{dr}{d\tau }=-\frac{\omega }{r\sqrt {1-({v}'/c)^2} }\hat {u}^r
\end{equation}
where $\hat {u}^r=\sqrt {g_{rr} } dr/d\tau =dr/d\tau $ is the physical 
velocity in the radial direction relative to the rotating frame.

The physical (measured in m/s$^{2})$ value for the tangential acceleration 
is
\begin{equation}
\label{eq29}
\hat {a}^\phi =\sqrt {g_{\phi \phi } } a^\phi =ra^\phi =\underbrace 
{-\frac{\omega }{\sqrt {1-({v}'/c)^2} }\hat {u}^r}_{Coriolis\,\,accel}.
\end{equation}
Again, this zeroth order (Newtonian) term arises solely because we have an 
off-diagonal first order time-space component in the metric, and is of this 
value solely because of the particular choice of simultaneity inherent in 
(\ref{eq21})(a).\footnote{ Note that Coriolis type acceleration does not generally 
arise with alternative synchronization schemes in translation (which also 
give rise to first order off diagonal time-space terms in the metric.) See 
(\ref{eq1}) and (\ref{eq20}). This is because, in such schemes, the off diagonal quantity, 
\textit{$\kappa $}, is a constant, and not a function of position. Thus, in the derivatives 
taken to find the Christoffel symbols $\Gamma ^{\alpha }_{\beta \gamma 
}$ , all such symbols equal zero, and no Coriolis type term results.}

\medskip
\noindent
\underline {Curvature}

Since the rotating frame metric was obtained via a transformation from the 
(flat 4D space) lab metric, Riemann curvature in both 4D frames is zero. For 
the 2D subspace of the rotating frame, it should be obvious from the 
\textit{r-$\phi $} submatrix of (\ref{eq22}) that curvature for that subspace is also zero. This will 
be discussed in more detail herein, but for the present, we note that this 
conclusion is conditional, in certain ways, upon our choice of simultaneity, 
inherent in (\ref{eq21})(a).

\medskip
\noindent 
\underline {\textbf{Conclusions:}}

\begin{enumerate}
\item The well known zeroth order (Newtonian) Coriolis acceleration found in rotation arises from first order off diagonal time-space terms in the metric, and those terms vary with choice of simultaneity.
\item For any AVS type (first order) synchronization choice having simultaneity other than that of the lab, application of general relativistic principles results in a predicted Coriolis acceleration different at zeroth order from that found in nature.
\item Any transformation differing from (\ref{eq21}) only in second order should produce the same zeroth order (Newtonian) Coriolis acceleration.
\end{enumerate}

\subsection{Coriolis Acceleration and Rotation via LCIFs }
\label{subsec:coriolis}
With the traditional LCIF approach to rotating frames, which employs LCIFs 
with Minkowski metrics, it does not appear that the Coriolis acceleration 
can be derived. To have Coriolis acceleration, there \textit{must} be off diagonal 
\textit{t-$\phi $} terms in the metric, such as those in (\ref{eq22}). But the traditional LCIF 
approach, due the imposition of Minkowski coordinates (Lorentz metric) in 
the LCIF used as a local surrogate for the rotating frame, does not have 
this characteristic. Thus, it fails as a complete analysis approach to 
rotating frames. In fact, it is demonstrably quite wrong.

Employing conventionality of synchronization (non-Lorentz metric), however, 
one can induce a metric in the LCIF that does indeed produce the correct 
Coriolis acceleration. With the respective line elements from (\ref{eq19}) and (\ref{eq22}), 
one can readily show that for the conventionality transformation (\ref{eq1}) from 
Einstein synchronization originally in the LCIF to the appropriate LCIF 
synchronization, we need $\kappa =v/c^{2}$ (with $C$ = 1), which from (\ref{eq19}), is 
seen to be a first order shift ($v/c)$ in synchronization for time. This, of 
course, means that synchronization is \textit{not} conventional for rotating frames, as 
there is \textit{only one synchronization scheme for which theory matches experiment}. 

Note that this particular synchronization choice is identical to shifting 
simultaneity in the LCIF to that of the lab, and thus to the choice taken in 
the transformation (\ref{eq21}). This is also equal to the choice advocated in 
Sec. \ref{subsubsec:mylabel1} herein, as necessary to keep time 
continuous and single valued.

\medskip
\noindent
\underline {\textbf{Conclusion}}\textbf{:} The only synchronization scheme 
for the LCIFs that provides complete and correct analysis of rotating 
systems (including Coriolis acceleration) has the same simultaneity as the 
lab. This is also the only such scheme that yields continuous, single valued 
time and clocks in synchronization with themselves in the rotating frame 
itself.

\section{Alternative Theories}
\label{sec:mylabel3}
We now summarize theories alternative to the traditional LCIF approach, 
which attempt to resolve the conundrums stated earlier herein. For thorough 
introduction to each, see Ref. \cite{Rizzi:2004} and citations therein.

\subsection{Selleri's Preferred Frame}
\label{subsec:selleri}
In his early work\cite{Selleri:2004}$^{,}$\cite{Selleri:1997}, Selleri 
used a number of arguments, including the Sagnac experiment, to posit the 
existence a preferred reference frame, and concomitantly, to an absolute 
simultaneity shared by all observers. He determined the transformation from 
the preferred frame to another inertial frame traveling at arbitrary speed 
$v$ with respect to the preferred frame that upholds these and other 
experimentally imposed requirements. Using an ``S'' subscript to refer to a 
non-preferred frame in the Selleri scheme, non-subscripted capital letters 
to designate the preferred frame, and a velocity directed along the $X$ axis, 
this transformation is
\begin{equation}
\label{eq30}
\begin{array}{l}
 t_S =\sqrt {1-v^2/c^2} 
\,T\,\,\,\,\,\,\,\,\,\,\,\,\,\,\,\,\,\,\,\,\,\,\,\,\,\,\,\,\,\,(a) \\ 
 x_S =\frac{1}{\sqrt {1-v^2/c^2} }\left( {X-vT} 
\right)\,\,\,\,\,\,\,\,\,\,\,\,\,(b) \\ 
 y_S =Y\quad \quad \quad \quad \quad \quad \quad \quad \quad \,\,\,\,\,(c) 
\\ 
 z_S =Z\quad \quad \quad \quad \quad \quad \quad \quad \quad 
\,\,\,\,\,(d)\,. \\ 
 \end{array}
\end{equation}
For future reference, and comparison with other approaches, we note that 
Selleri's metric is non-time-orthogonal, and has the form
\begin{equation}
\label{eq31}
g_{\mu \nu } =\left[ {{\begin{array}{*{20}c}
 {-1} \hfill & 0 \hfill & {v/c} \hfill & 0 \hfill \\
 0 \hfill & 1 \hfill & 0 \hfill & 0 \hfill \\
 {v/c} \hfill & 0 \hfill & {1-v^2/c^2} \hfill & 0 \hfill \\
 0 \hfill & 0 \hfill & 0 \hfill & 1 \hfill \\
\end{array} }} \right].
\end{equation}
For reasons of consistency in comparison with other metrics, we have moved 
the components arising from the $x_{S}$ coordinate of (\ref{eq30}) from the second 
row and column of (\ref{eq31}) to the third.

Note, particularly, that in Selleri's transformation, like the 
conventionality transformations discussed in Sec. 
\ref{sec:alternative}, the coordinates $t_{S}$ and $x_{S}$ represent 
values measured with physical standard clocks and meter sticks, 
respectively. But, again like the transformations of Sec. 
\ref{sec:alternative}, $x_{S}$ does \textit{not} represent proper length, i.e., it 
does \textit{not} represent 4D physical length, the 4D interval length, along the 
$x_{S}$ axis. See Fig. 7, where along the $x_{S}$ axis an observer fixed to 
that axis would lay down 3.3 meter sticks to cover a 4D physical length of 
3.0 meters.

\begin{figure}
\centerline{\includegraphics{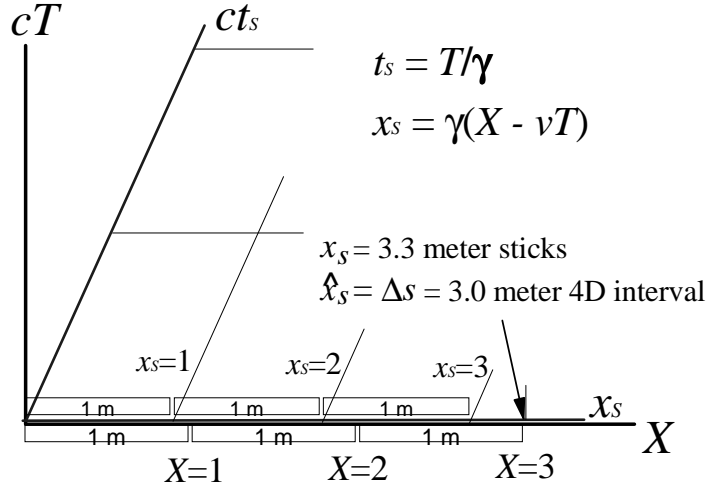}}
\caption{. Selleri Transformation}
\label{fig7}
\end{figure}

This means that observers in both frames would consider meter sticks fixed 
in the $S$ frame to measure $C \ne $ 2$\pi r$.

As Selleri showed, within the S system, one-way speed of light is 
anisotropic, but the two-way speed remains isotropic and equal to $c$. As many 
have pointed out\cite{Anderson:1998}, distant clock synchronization and 
one-way light speed are interdependent, so no absolute determination of 
one-way speed can be made, and thus Selleri's transformation appears 
consistent with all possible real-world measurements.

Selleri proposed his transformation for both translation and rotation. But 
he considered the Sagnac experiment, with its ostensible difference in 
one-way light speeds, to be a key pillar supporting it. In that test, the 
distant clock is actually the same as the original clock, and reasonable 
arguments can be made that it provides a valid measurement of one-way speed.

Note that for application to rotation, with reference to cylindrical 
coordinates such as those in (\ref{eq21}), $v = \omega r$, and the $x_{s}$ direction in 
(\ref{eq30}) is taken as the circumferential, or \textit{$\phi $}, direction, though it is a measure 
of distance and not angle. The reader should be aware that, in this case, 
(\ref{eq31}) is not a true rotating frame metric because it is only applicable, as 
written, along a single circumferential line. For applicability to the 
entire rotating frame, the metric would need an $r$ dependence, and would be 
significantly more complicated than (\ref{eq31}) (See Sec. 
\ref{subsubsec:mylabel2}.).

Selleri has modified his position recently, and we review this development 
in Sec. \ref{subsec:synchronization}.

\medskip
\noindent
\underline {\textbf{To summarize}}:

\medskip
\noindent
\underline {Michelson-Morley}

Selleri's approach predicts a null MM signal.

\noindent 
\underline {Lorentz Contraction}

Selleri predicts circumferential Lorentz contraction effects in the rotating 
frame.

\noindent 
\underline {Time Gap and Coriolis}

Because Selleri's simultaneity (see (\ref{eq30}), first line) equals that of the 
lab, there is no discontinuity in time such as that shown in Fig. 1. If 
his metric (\ref{eq31}) could be recast as a global metric (with $r$ dependence), like 
that of (\ref{eq22}), his approach would predict a Coriolis acceleration in 
agreement with the low speed value.

\noindent
\underline {Other Issues Raised by Selleri}

It should be mentioned that Selleri has raised other challenges to accepted 
tenets of SRT, including starlight aberration in inertial reference frames 
and certain aspects of relativistic acceleration. These deserve serious 
consideration, but will not be addressed herein.

\subsection{Rizzi, Ruggiero, and Serafini's Desynchronization}
\label{subsec:rizzi}
Rizzi, Ruggiero, and 
Serafini\cite{Rizzi:2005}$^{,}$\cite{Rizzi:2006}$^{,}$\cite{Rizzi:2002} 
(RRS) have noted that slow (relative to the rotating disk) transport of 
standard clocks 360$^{o}$ around the circumference of a rotating disk leads 
to a time difference between the traveled clock and the clock left at the 
starting point.

This desynchronization effect has, in fact, been shown by Anderson et 
al\cite{Anderson:1998} to be generally characteristic of slowly 
transported clocks in any non-Einstein synched coordinate system. That is, 
unless the 4D coordinates employ Einstein synchronization, slow transport of 
any standard clock will yield a time difference (a desynchronization) 
between that clock and the standard clock at the destination point. For 
Einstein synchronization, the traveled and destination clocks are in synch.

So if one assumes Einstein synchronization around the rim of the rotating 
disk, then a slow clock transport around that rim results in no 
desynchronization between the destination clock (at 360$^{o})$ and the 
traveled clock. However the clock at 360$^{o}$ is out of synchronization 
with the clock at 0$^{o}$ (because they have been Einstein synched -- see 
time gap of Fig. 1), and thus the traveled clock is desynchronized with 
the original starting point clock. Quantitatively, the amount of the 
desynchronization equals the time gap of Fig. 1.

For clocks traveling slowly in the cw and ccw directions, the difference 
between those two clocks after 360$^{o}$ equals the Sagnac time difference 
(\ref{eq4}). That is, both clocks are ``desynchronized'' from the clock left at the 
starting point, and the amount of this desynchronization difference turns 
out to be equal to the Sagnac difference. For these authors, this is the 
``physical root'' of the Sagnac effect, and they thus suggest that this 
resolves the issue.

However, the coinciding of slow moving clock desynchronization, the Sagnac 
time difference, and the time gap of Fig. 1 (including the cw equivalent 
of Fig. 1), do not, I submit, resolve the inconsistency issues of Sec. 
\ref{subsec:internal} (time discontinuity, limit case, Lorentz 
contraction) or the Coriolis issue with LCIFs of Sec. 
\ref{subsec:coriolis}. Further, even for the limited issue involving 
Sagnac, one could make other arguments for why these three phenomena 
converge quantitatively.

The transformation used in desynchronization is twofold. First, one 
transforms using the Lorentz transformation from the lab to the LCIF frame, 
which is assumed via the principle of locality to reflect the rotating frame 
itself (locally.) Second, if one chooses, one can resynchronize the LCIF 
(the rotating frame) via the resynchronization transformation (\ref{eq19}) and 
obtain the metric (\ref{eq20}), repeated here for convenience,
\begin{equation}
\label{eq32}
g_{\alpha \beta } =\left[ {{\begin{array}{*{20}c}
 {-1} \hfill & 0 \hfill & \kappa \hfill & 0 \hfill \\
 0 \hfill & 1 \hfill & 0 \hfill & 0 \hfill \\
 \kappa \hfill & 0 \hfill & {1-\kappa ^2} \hfill & 0 \hfill \\
 0 \hfill & 0 \hfill & 0 \hfill & 1 \hfill \\
\end{array} }} \right].
\end{equation}
In (\ref{eq32}), the resynchronization direction is oriented along a circumference 
and similar to the Selleri case is a distance, not angle, measure. For zero 
resynchronization from Einstein synchronization, \textit{$\kappa $} = 0, and (\ref{eq32}) reduces to 
the Lorentz (or Minkowski) metric. As RRS have shown, regardless of the 
synchronization chosen (regardless of the form of the metric), the time on a 
slow moving (with respect to the rotating frame) clock will always turn out 
to be the same value, i.e., it is invariant with respect to 
resynchronization transformations.

\medskip
\noindent
\underline {\textbf{To summarize}}:

\medskip
\noindent 
\underline {Michelson-Morley}

The RRS desynchronization approach predicts a null MM signal.

\noindent
\underline {Lorentz Contraction}

RRS appear to accept circumferential Lorentz contraction, deduced via the 
traditional logic, which is challenged herein in Secs. 
\ref{subsubsec:lorentz} and 6.

\noindent
\underline {Time Gap and Coriolis}

RRS advocate conventionality of simultaneity in the rotating frame and thus 
will have a time discontinuity that varies with choice of simultaneity. For 
the one choice of simultaneity equal that of the lab, there would be no such 
discontinuity and clocks would be in synchronization with themselves.

The RRS desynchronization approach will yield different Coriolis type 
accelerations for different choices of simultaneity. Only for the continuous 
time choice will it equal the classical, low speed value.

\subsection{Bel's Optical and Mechanical Metrics}
\label{subsec:mylabel1}
Bel\cite{Bel:2004}$^{,}$\cite{Bel:1}$^{,}$\cite{Bel:2000}$^{,}$\cite{Ref:3}, 
along with colleagues, contended that two metrics exist, one optically 
based, and one mechanically based. In a rotating frame these would turn out 
to be different, and this difference would manifest as a non-null 
Michelson-Morley result.

Quantitatively, this non-null signal would be quite close to the anomalous 
signal measured by Brillet and Hall. Bel notes\cite{Ref:4} that according 
to his analysis, in the earth fixed frame, this signal must be 13$^{o}$ from 
the E-W direction, but regrettably, the phase angle of the Brillet-Hall 
signal was not recorded. Had it been found to be in the direction predicted, 
it would have lent strong support to Bel's theory. Of course, a direction 
other than this would disprove his theory.

Likewise, and similar to Klauber's approach (to be described below), a null 
rotating frame MM signal (which would show Brillet-Hall's signal truly was 
spurious) would also negate Bel's approach.

\medskip
\noindent 
\underline {\textbf{To summarize}}:

\medskip
\noindent 
\underline {Michelson-Morley Experiment}

The Bel et al approach predicts a non-null MM signal close to that found by 
Brillet and Hall.

\noindent
\underline {Byl et al Experiment}

The Byl et al\cite{Byl:1} experiment, a presumed first order test of SRT, 
entailed a one way propagation of light through air and glass. I believed 
their null result disproved Bel's approach, but was informed by Bel that it 
does not, though I am not convinced, as I have not seen an analysis 
supporting this contention.

\noindent
\underline {Lorentz Contraction, Time Gap, and Coriolis}

I have not investigated these issues with regard to the Bel et al approach.

\subsection{Nikolic's Local Proper Coordinate Frames}
\label{subsec:nikolic}
Nikolic\cite{Nikolic:1} analyzes rotating frames by employing a different 
local frame with proper coordinates for each rotating frame observer. 
Adopting terminology from Misner, Thorne, and Wheeler\footnote{ See Sec. 
13.6 in Ref. \cite{Misner:1973}. Nikolic has not invented proper 
coordinate frames, just applied them to rotation.}, he presently calls these 
``proper coordinate frames'', although in past work, he deemed them ``Fermi 
frames''. Actual measurements made in these local frames would equal 
coordinate values, since coordinate values are purposely chosen to equal 
proper values.

Nikolic believes the rotating frame cannot be correctly described by a 
single coordinate system, but only by an infinite set of local proper 
coordinate frames. He states ``even if there is no relative motion between 
two observers [fixed on a rotating disk, for example], they belong to 
different frames if they do not have the same position.''\cite{Ref:5} 
Further, he considers this generally true of all non-inertial systems, ``two 
observers at different positions but with zero relative velocity may be 
regarded as belonging to the same coordinate frame only if they move 
inertially in flat space-time.''

The reader must read Nikolic directly to gain full appreciation of his 
justification for this seemingly strange position. We do note that, given 
this position, he may appear to resolve certain conundrums involving 
relativistic rotation. For example, if there can be no global frame, then 
one might argue there is no global time gap issue. However, some, including 
myself, may not consider this a resolution, as all measurements made in the 
physical world are of necessity global, and to make predictions, one must 
integrate over adjacent local frames (as in the traditional LCIF approach). 
Carrying out such integration for time, Nikolic's local frames yield a time 
gap such as that shown in Fig. 1.

Nikolic also presents interesting perspectives on length contraction and 
clock rates. He argues that ``one can study the relativistic contraction in 
the same way as in the conventional approach with Lorentz frames''. (See 
Secs. \ref{subsubsec:lorentz} and 6 
herein for reasons why this may not be reasonable.) He also deduces the time 
seen by an observer on the rotating frame, which is seen to oscillate with 
each rotation, and only when averaged over a whole rotation, does it equal 
the traditional Lorentz dilated value.

Nikolic's metric, expressed in \textit{local}, not global coordinates, with non-rotation 
induced accelerations taken as zero and Cartesian coordinates transformed to 
cylindrical coordinates, is
\begin{equation}
\label{eq33}
g_{\mu \nu } =\left[ {{\begin{array}{*{20}c}
 {-\left( {1-\left( {{\omega }'{r}'/c} \right)^2} \right)} \hfill & 0 \hfill 
& {\left( {{\omega }'{r}'/c} \right){r}'} \hfill & 0 \hfill \\
 0 \hfill & 1 \hfill & 0 \hfill & 0 \hfill \\
 {\left( {{\omega }'{r}'/c} \right){r}'} \hfill & 0 \hfill & {\left( {{r}'} 
\right)^2} \hfill & 0 \hfill \\
 0 \hfill & 0 \hfill & 0 \hfill & 1 \hfill \\
\end{array} }} \right]
\end{equation}
where primes indicate values in the local coordinate system. The presence of 
the extra $r^{/}$ in the off diagonal components of (\ref{eq33}) (compared to (\ref{eq31}), 
for example) turns the angular $d\phi $ term in the position vector to the 
distance \textit{rd$\phi $} term in the line element.

Note that at the origin of each local frame, $r^{/}$ = 0, and (\ref{eq33}) becomes 
the time-orthogonal Lorentz metric (effectively, for cylindrical 
coordinates). As one moves away from the local origin, the coordinates 
become more and more non-time-orthogonal. Thus, a point slightly removed 
from the origin of the local frame would have a different metric from that 
of the local frame with its origin at that point.

From the analysis of Sec. \ref{subsubsec:rotation}, one should be 
able to see that the derivatives in the calculation of the Christoffel 
symbols manifest acceptable Coriolis and centrifugal acceleration terms. 
This is true even though, at the local origin, the metric is diagonal and 
Lorentzian, similar in this regard to the traditional LCIF approach (which 
does not yield Coriolis acceleration.)

\medskip
\noindent
\underline {\textbf{To summarize}}:

\medskip
\noindent 
\underline {Michelson-Morley}

Nikolic's approach predicts a null MM signal.

\noindent
\underline {Lorentz Contraction}

Nikolic seems to argue that the traditional logic for Lorentz contraction 
can be applied.

\noindent 
\underline {Time Gap and Coriolis}

Nikolic shows a time discontinuity globally, if one integrates his local 
frame time values around the entire circumference. His metric does give rise 
to the correct low speed Coriolis acceleration, and to centrifugal 
acceleration.

\subsection{Klauber's NTO Approach}
\label{subsec:klauber}
Klauber\cite{Klauber:2004}$^{,}$\cite{Klauber:2005}$^{,}$\cite{Klauber:2} 
analyzed relativistic rotation using a fully differential geometric 
approach. He started with the widely used transformation to a rotating frame 
(\ref{eq21}), i.e.,
\begin{equation}
\label{eq34}
\begin{array}{l}
 ct=cT\quad \quad \quad \quad \quad \mbox{(a)} \\ 
 r=R\quad \quad \quad \quad \;\;\;\quad \mbox{(b)} \\ 
 \phi =\Phi -\omega T\quad \quad \quad \;\,\mbox{(c)} \\ 
 z=Z\quad \quad \quad \quad \quad \;\;\;\mbox{(d)}\,\mbox{,} \\ 
 \end{array}
\end{equation}
which yields the non-time-orthogonal (NTO) rotating frame metric (\ref{eq22}), i.e.,
\begin{equation}
\label{eq35}
g_{\alpha \beta } =\left[ {{\begin{array}{*{20}c}
 {-(1-\textstyle{{r^2\omega ^2} \over {c^2}})} \hfill & 0 \hfill & 
{\textstyle{{r^2\omega } \over c}} \hfill & 0 \hfill \\
 0 \hfill & 1 \hfill & 0 \hfill & 0 \hfill \\
 {\textstyle{{r^2\omega } \over c}} \hfill & 0 \hfill & {r^2} \hfill & 0 
\hfill \\
 0 \hfill & 0 \hfill & 0 \hfill & 1 \hfill \\
\end{array} }} \right].
\end{equation}
He then used tensor analysis with that metric to deduce relevant properties 
of rotating frames.

To find light speeds, Klauber set the line element for \textit{ds} found from (\ref{eq35}) 
equal to zero for two conditions: 1) circumferential direction (\textit{dr = dz = }0), and 2) 
radial direction ($d\phi $ = \textit{dz} = 0.) These equations could then be solved for 
1) \textit{d$\phi $}/\textit{dt} and 2) \textit{dr/dt}.

He then assumed the standard differential geometry approach for finding 
physical components (those measured with actual instruments according to the 
theory) applied and found the one-way physical velocity (see (\ref{eq14})) in the 
circumferential direction to be
\begin{equation}
\label{eq36}
v_{light,phys,circum} \;\;=\;\;\frac{\sqrt {g_{\phi \phi } } d\phi }{\sqrt 
{-g_{tt} } dt}\;\;=\;\;\frac{-\;\omega r\;\pm \;c}{\sqrt 
{1-\textstyle{{\omega ^2r^2} \over {c^2}}} }=\;\;\frac{-\;v\;\pm \;c}{\sqrt 
{1-\textstyle{{v^2} \over {c^2}}} },
\end{equation}
and the one-way physical radial direction velocity to be
\begin{equation}
\label{eq37}
v_{light,phys,radial} =\frac{d\hat {r}}{d\hat {t}}=\frac{\sqrt {g_{rr} } 
dr}{\sqrt {-g_{tt} } dt}=c.
\end{equation}
Using (\ref{eq36}) and (\ref{eq37}) in the standard MM analysis\cite{Klauber:2005}, 
Klauber found a time delay in the circumferential direction over that in the 
radial (and $z$ direction) of $\raise.5ex\hbox{$\scriptstyle 1$}\kern-.1em/ 
\kern-.15em\lower.25ex\hbox{$\scriptstyle 2$} v^{2}$/$c^{2}$, and thus an 
expected non-null MM signal very close to that found by Brillet and Hall.

Klauber also reasoned that the method for finding physical components from 
differential geometry should allow ready determination of whether or not 
Lorentz contraction exists in a rotating frame. Thus,
\begin{equation}
\label{eq38}
\begin{array}{l}

\mbox{measured}\,\,\mbox{length}\,\,\mbox{in}\,\,\mbox{rotating}\,\,\mbox{frame}\,\,\mbox{=}\,\,d\phi 
_{phys} \\ 
 =d\hat {\phi }=\sqrt {g_{\phi \phi } } d\phi =rd\phi =r(\phi _2 -\phi _1 ) 
\\ 
 \end{array},
\end{equation}
which from (\ref{eq34}), with endpoint measurements simultaneous (i.e., $t_{1}$ = 
$t_{2} \to $  $T_{1}=T_{2})$, equals
\begin{equation}
\label{eq39}
\begin{array}{l}
 R\left( {\Phi _2 +\omega {\kern 1pt}{\rm T}_2 -\Phi _1 -\omega {\kern 
1pt}{\rm T}_1 } \right)=R\left( {\Phi _2 -\Phi _1 } \right) \\ 
 =\,\mbox{measured}\,\,\mbox{length}\,\,\mbox{in}\,\,\mbox{lab} \\ 
 \end{array}.
\end{equation}
Thus, if the measured length between two 3D points is the same in both the 
rotating frame and the lab, there is no Lorentz contraction. Note that, via 
this logic, we would need a metric with $g_{\phi \phi } \ne r^{2}$ to 
have Lorentz contraction.

\medskip
\noindent 
\underline {\textbf{To summarize}}:

\medskip
\noindent 
\underline {Michelson-Morley Experiment}

Using differential geometry with physical component determination, Klauber 
predicted a non-null MM signal close to that found by Brillet and Hall.

\noindent
\underline {Byl et al}

At first perusal, one might consider Klauber's approach disproved by the Byl 
et al\cite{Byl:1} result, but a thorough 
analysis\cite{Klauber:2000} showed this not to be the case.

\noindent 
\underline {Lorentz Contraction}

Using the same differential geometric approach, Klauber predicted no Lorentz 
contraction and $C$ = 2$\pi r$.

\noindent 
\underline {Time Gap and Coriolis}

Similar to Selleri, simultaneity in the rotating frame and the lab are the 
same, so there is no discontinuity in time such as that shown in Fig. 1. 
Klauber contends there is only one correct simultaneity in rotation, the one 
for which no time discontinuity exists. Coriolis acceleration determination 
is correct in the low speed limit, as shown in Sec. 
\ref{subsubsec:rotation}. Centrifugal acceleration is also found 
correctly.

\noindent 
\underline {Analysis Depends on Two Things:}

All of Klauber's predictions depend on two things: 1) correctness of 
transformation (\ref{eq34}) (i.e., is this the actual transformation nature has 
chosen?) and 2) physical lengths as determined using differential geometry 
equal those actually measured with real world meter sticks.

\section{Convergence of Seemingly Disparate Theories}
\label{sec:convergence}
\subsection{Synchronization Conventionality Means Selleri = SRT for Inertial Frames}
\label{subsec:synchronization}
In an enlightening article\cite{Rizzi:2005}, RRS show that, accepting 
conventionality of synchronization as true, Selleri coordinates in any 
non-preferred Selleri inertial frame can be converted into Minkowski 
coordinates (i.e., the frame is actually a Lorentz frame). If the 
non-preferred Selleri frame has velocity $v$ relative to the preferred frame, 
then resynchronization with \textit{$\kappa $} = $v/c$, i.e.,
\begin{equation}
\label{eq40}
ct_M =ct_S -(v/c)x_S \quad ,
\end{equation}
will transform the Selleri coordinates to Minkowski coordinates. 
Graphically, this is illustrated in Fig. 8, where the non-time-orthogonal 
Selleri coordinates \textit{ct}$_{S}-x_{S}$ are shown converted to orthogonal 
(Minkowski) coordinates \textit{ct}$_{M}-x_{M}$.

\begin{figure}
\centerline{\includegraphics{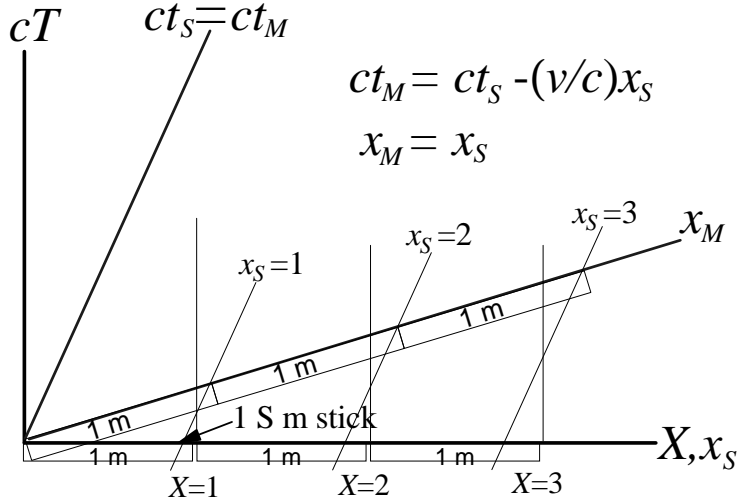}}
\caption{. Synch Transformation for Selleri Coords}
\label{fig8}
\end{figure}

Thus, any inertial frame proposed by Selleri as non-preferred can in reality 
be chosen as his preferred frame (i.e., the frame with isotropic one-way 
light speed.) And thus, there can be no true preferred frame. 
Selleri\cite{Selleri:2005} now agrees with this, though he has some 
difference with RRS on philosophic and interpretational levels.

RRS then extend this logic to rotating frames, and surmise that the 
difference between Selleri's approach and the traditional LCIF approach is 
simply a choice of simultaneity, and that any such choice works, though some 
may be more convenient for certain problems than others. From this point of 
view, they consider Selleri's limit case speed of light paradox (Sec. 
\ref{subsubsec:selleri}) to be no paradox at all, but merely arises 
due to differing choices for simultaneity in the rotating and LCIF frames.

While I agree wholeheartedly with the analysis by RRS with regard to 
inertial frames, I disagree with regard to rotating frames. For reasons 
delineated herein, I consider Selleri's choice of simultaneity to be the 
only feasible one on a rotating platform.

\medskip
\noindent
\underline {\textbf{Conclusion:}}

Any inertial frame originally proposed by Selleri as non-preferred can be 
shown, via a shift in simultaneity, to be a Lorentz frame, and thus no 
unique preferred frame actually exists. RRS consider this result applicable 
to non-inertial, and in particular rotating, frames, and thus contend that 
Selleri's approach is essentially equivalent to their own (LCIFs with 
conventionality of simultaneity.)

\subsection{Klauber Similar to Selleri for Rotation}
\subsubsection{Time for Selleri and Klauber}
Selleri's original approach, as applied to rotation, and Klauber's approach 
share the principle that simultaneity in the rotating frame is the same as 
that in the lab. Clocks run at different rates in the rotating frame, but 
observers in both frames can agree that zero time passes on their clocks 
between two spatially separate events.

The rates on physical standard clocks are also the same for both approaches. 
The difference is only in the definition of coordinate time. For Selleri, 
using (\ref{eq31}) and (\ref{eq30}), a standard clock fixed in the rotating frame has
\begin{equation}
\label{eq41}
\begin{array}{c}
 \mbox{Selleri}\,\mbox{standard}\mbox{ clock}\,\mbox{time}\,=d\hat {t}_S 
=\sqrt {-g_{S\,tt} dt_S dt_S } \\ 
 =\sqrt {-(-1)dt_S ^2} =dt_S =\sqrt {1-(\omega r)^2/c^2} dT\,, \\ 
 \end{array}
\end{equation}
and his coordinate time equals the time on a standard clock at a given 
location.

For Klauber, coordinate time corresponds to time on a clock located at the 
center of rotation. From (\ref{eq35}),
\begin{equation}
\label{eq42}
\begin{array}{c}
 \mbox{Klauber}\,\mbox{standard}\mbox{ clock}\,\mbox{time}\,=d\hat {t}_K 
=\sqrt {-g_{K\,tt} dt_K dt_K } \\ 
 =\sqrt {-(-1+(\omega r)^2/c^2)dt_K ^2} =\sqrt {1-(\omega r)^2/c^2} dT\,, \\ 
 \end{array}
\end{equation}
and thus, (\ref{eq41}) equals (\ref{eq42}). 

\medskip
\noindent
\underline {\textbf{Conclusion:}} The Selleri and Klauber approaches, in 
terms of predictions for physically measured time, are essentially 
equivalent. Both have the same simultaneity, degree of 
non-time-orthogonality (slope of time axis), and time dilation.

\subsubsection{Space for Selleri and Klauber}
\label{subsubsec:space}
With regard to distances measured with standard (physical) meter sticks, 
however, the definition of the spatial coordinate in the circumferential 
direction becomes critical. As shown in Sec. 
\ref{subsec:selleri}, and illustrated in Fig. 7 therein, Selleri 
defines his $x_{S}$ coordinate (in similar fashion to his choice of time 
coordinate) as equal to the value measured with actual physical meter 
sticks. Via his transformation (\ref{eq30})(b), the Lorentz contraction of those 
meter sticks is then ``built in''. Given the definition of $x_{S}$, and 
Selleri's chosen transformation, Lorentz contraction in the rotating frame, 
agreed to by all observers in all frames, must result.

We note that this choice of transformation is built on a prejudice that we 
must have Lorentz contraction in the rotating frame. And this prejudice is 
based on the traditional, widely dispersed logic, which, is suggested in 
Secs. \ref{subsubsec:mylabel1}, \ref{subsubsec:lorentz} and 
6, to be inconsistent. Lorentz contraction in 
rotation may exist, but there is no sound theoretical reason to insist, \textit{a priori}, 
that it must. Only experiment can answer this question unequivocally.

With regard to space, Klauber and Selleri have two differences, one trivial, 
one not so. Trivially, Klauber uses cylindrical coordinates with \textit{d$\phi $} in the 
circumferential direction, whereas Selleri uses orthogonal coordinates and 
\textit{dx}$_{s}$ in the circumferential direction.

Non-trivially, Selleri takes his coordinate value \textit{dx}$_{s}$ as equal to the 
number of real world meter sticks (and not \textit{ds}), whereas Klauber, in the usual 
differential geometry fashion, considers \textit{ds} to be the distance measured in 
real world meter sticks, with \textit{d$\phi $} merely a coordinate value. Thus, we have
\begin{equation}
\label{eq43}
\begin{array}{l}
 \mbox{Selleri}\,\mbox{interval}\,\,=ds=\sqrt {1-\omega ^2r^2/c^2} dx_s \\ 

\mbox{Selleri}\,\mbox{real}\,\mbox{world}\,\mbox{meter}\,\mbox{sticks}\,\,=dx_s 
\,\,\left( {\ne ds} \right)\, \\ 
 dx_s =\frac{1}{\sqrt {1-\omega ^2r^2/c^2} 
}dX=\mbox{Lorentz}\,\mbox{contracted}\,\mbox{measure,} \\ 
 \end{array}
\end{equation}
and
\begin{equation}
\label{eq44}
\begin{array}{l}
 \mbox{Klauber}\,\mbox{interval}\,\,=ds\,\,\,\,\,\left( 
{\mbox{same}\,\mbox{as}\,\mbox{Selleri}} \right) \\ 

\mbox{Klauber}\,\mbox{real}\,\mbox{world}\,\mbox{meter}\,\mbox{sticks}\,\,=ds=\sqrt 
{g_{\phi \phi } } d\phi =rd\phi \\ 
 ds=Rd\Phi 
\,=\,\mbox{no}\,\mbox{Lorentz}\,\mbox{contraction}\,\mbox{measurement.} \\ 
 \end{array}
\end{equation}
Note that if Klauber were to assume at the outset, as Selleri did, that 
Lorentz contraction of meter sticks must exist, then one would simply, and 
alternatively, reinterpret \textit{ds =} \textit{rd$\phi $} as a number not equal to the number of real 
world meter sticks used to measure distance. That is, as
\begin{equation}
\label{eq45}
rd\phi =\sqrt {1-\omega ^2r^2/c^2} \underbrace {dx_K }_{\begin{array}{l}
 \mbox{meter}\,\mbox{sticks} \\ 
 \mbox{for}\,\mbox{alternative} \\ 
 \mbox{Klauber} \\ 
 \end{array}}=\sqrt {1-\omega ^2r^2/c^2} \underbrace {dx_S 
}_{\begin{array}{l}
 \mbox{meter}\,\mbox{sticks} \\ 
 \mbox{for}\,\mbox{Selleri} \\ 
 \end{array}}.
\end{equation}
Thus, the difference between Selleri and Klauber, with regard to spatial 
measurements, is solely one of interpretation. If one insists on Lorentz 
contraction (because presumably it has been found experimentally in 
rotation), then it is merely an issue of reinterpreting Klauber's 
coordinates in terms of real world meter sticks, which are \textit{a priori} assumed to 
contract.

\medskip
\noindent
\underline {\textbf{Conclusions:}} 

\begin{enumerate}
\item The Selleri and Klauber approaches, in terms of predictions for real world measured lengths, can be considered essentially equivalent, if one assumes ahead of time that Lorentz contraction of real world meter sticks exists. Any difference then lies solely in interpretation of coordinate values.
\item Klauber originally used the standard differential geometry interpretation (for purely spatial spaces) of the interval \textit{ds} as equal to the number of real world meter sticks. For this interpretation, there is no Lorentz contraction.
\end{enumerate}
\subsection{Are All Approaches Equivalent?}
If the Selleri approach, given conventionality of synchronization, is 
equivalent to the traditional, or RRS, approach, and the Klauber approach is 
equivalent to the original Selleri approach, then one could ask if all 
approaches converge to the same theory. There are two aspects to the answer.

First, it does not appear that the Bel or Nikolic approaches can be 
enveloped into the fold. They seem distinctly different from the other 
three.

Second, if there is only one possible simultaneity in rotation, as is argued 
herein, then the Selleri approach is not equivalent to the RRS approach. The 
original Selleri approach and the Klauber approach would remain equivalent, 
varying only in the interpretation of coordinate values with regard to real 
world measurements of Lorentz contraction, if any.

\medskip
\noindent
\underline {\textbf{Conclusion:}} With regard to time and space, the 
(original) Selleri and Klauber approaches are equivalent (variation can 
occur in interpretation.) Selleri's approach can be extended to be 
equivalent to RRS only if simultaneity in rotation is conventional.

\section{A Complete Consistent Theory of Rotation?}
\label{sec:mylabel4}
\subsection{Need for Global Metric and Unique Time in a Good Theory}
\label{subsec:mylabel2}
The standard general relativistic approach can be summarized as the 
following.

Lay out a 4D coordinate grid, i.e., a 3D coordinate grid plus coordinate 
clocks at each point in the 3D grid. Using standard real world meter sticks 
and real world clocks at each 3D point, make measurements and determine the 
metric everywhere for the 4D grid chosen. (Note that the simultaneity scheme 
to be used is inherent in the clock settings chosen, and once the settings 
are made, every event has a unique time associated with it.)

This metric is then used with all laws of physics, generalized covariant 
laws - kinematic, dynamic, electromagnetic, gravitational - to deduce all 
observed phenomena. Additionally, the global geometry can be found by using 
the metric to determine the Riemann tensor.

Thus, a good and complete theory has

\begin{enumerate}
\item a global metric, and
\item a unique time at each event (unique clock at each 3D point).
\end{enumerate}
Note that 2) above is contrary to approaches to rotation that allow more 
than one time to be assigned to a given event, for a single specific 
synchronization scheme.

\subsection{A Good Global Metric for Rotation Must Exist}
Thus, there must exist a good global theory for rotation, as we can 
certainly climb aboard a rotating disk with standard meter sticks and clocks 
in hand, lay out a grid on the disk and have a metric which can be used to 
deduce all phenomena. Without such a coordinate grid and metric, it would 
not be easy, for example, to solve for the trajectory of a particle (free or 
not) as seen by an observer on a rotating disk. If we have a global 
coordinate system, we can specify the position coordinates as functions of 
time and have a very compact description, of enormous practical value, of 
the particle motion. How does one do that if the particle is crossing into a 
new coordinate system (infinitesimal, local) at every instant? It can 
probably be done, but the work doing it, and the final expression(s) showing 
it, are complex to extraordinary degree.

Thus, I claim that the best (i.e., the simplest, most straightforward) 
theory of rotation is one with a global metric inherent within that theory. 
The overriding question then should be ``what is a valid form for that 
metric?''

Note that this metric, if we are to obey the usual rules of GR physics, must 
have continuous, unique time values throughout. Note also, that the 
traditional approach of local LCIFs has not been shown to provide a global 
metric, suitable for solving all rotating frame problems.

\section{2005 Experiments}
\label{sec:mylabel5}
Three experiments expected to be crucial in deciding between 
Bel/Klauber type theories, which predicted non-null MM rotating frame 
signals, and others, were reported in 2005. However, with respect to the 
issue addressed herein of culling incorrect theories of rotation from the 
extant mix, two of these experiments, as originally reported, did not 
provide an answer to this question.

Each of the three experiments utilized a rotating apparatus and was a 
variation on the standard MM experiment. Their sensitivities were more than 
sufficient to detect any non-null signal due to the earth surface speed 
about its own axis, which would be of the order 10$^{-13}$. However, the 
primary focus in each experiment was to determine cosmic anisotropies in 
light speed (i.e. relative to the fixed stars, not the earth surface), and 
the method used to analyze this can completely obscure any local effect from 
the earth surface speed due to its rotation.

\begin{figure}
\centerline{\includegraphics{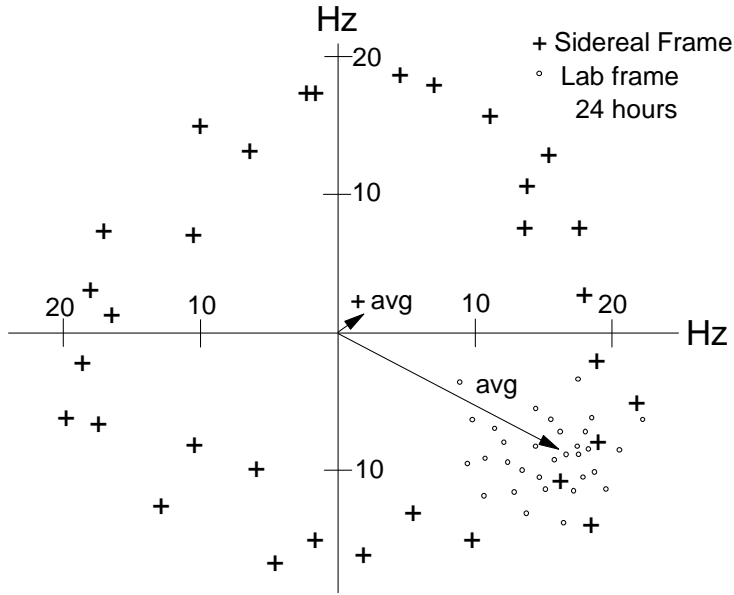}}
\caption{. Expected Data for Rotation Anisotropy}
\label{fig9}
\end{figure}

To see this, consider Fig. 9, which displays the type of data (but not 
the actual data) found by Brillet and Hall\cite{Brillet:1979}, and 
illustrated in Fig. 2 of their report. Data is taken over one rotation of 
the earth (24 hours) and can be plotted in two ways, using ``+'' symbols as 
relative to the fixed stars, and using ``o'' symbols relative to the earth's 
surface. Averaging the former nets an effectively null signal, i.e., zero 
cosmic anisotropy. Averaging the latter nets a significant non-null signal, 
suggestive of local anisotropy. This is the type of 
signal that occurred in the Brillet and Hall data, 
and though considered anomalous by most, was cited by Bel and 
Klauber as a possible indication of true anisotropy in the rotating earth 
frame. The magnitude of this local signal found by Brillet and Hall was 
remarkably close to what one would expect from a pre-relativistic analysis 
using the surface speed of the earth about its axis as the ``ether'' speed.

However, as we will see below, data from at least two of the 2005 tests 
showed something different.

\subsection{Antonini et al}
Antonini et al\cite{Antonini:2005} compared the resonance frequencies of 
two perpendicular optical resonators as their apparatus rotated relative to 
the lab. They took data samples at 1 second intervals for 76 hours. They 
first analyzed their data as a function of one resonator's axis relative to 
the south direction to obtain amplitudes for each individual rotation of 
their apparatus.

This is precisely the type of analysis needed for our purpose, and it showed 
variations on the order of 10$^{-14}$, approximately an order of magnitude 
smaller than the anomalous Brillet and Hall signal and thus, that required 
by Bel and Klauber.

Antonini et al then plotted the single rotation signal amplitudes against 
hours of test duration to discern any cosmic anisotropy (which would 
manifest as changes over several hours as the earth alignment with the 
heavens changed), and found no evidence for it.

\medskip
\noindent
\underline {\textbf{Conclusion}}\textbf{:} Antonini et al analyzed data in 
the earth fixed frame and found no signal such as that of Brillet and Hall, 
which might indicate rotating frame anisotropy of light speed.

\subsection{Stanwix et al}
Stanwix et al\cite{Stanwix:2005} rotated two orthogonally oriented 
cryogenic resonator-oscillators operating in whispering gallery modes to 
test for violations of Lorentz invariance. They found nothing to suggest 
light speed anisotropies, but unfortunately, for our purpose, the original 
analysis of their data was carried out entirely in the non-rotating earth 
centered frame. The computer driven analysis converted all data to this 
frame before analysis, and thus the results are meaningless with regard to 
determination of LLI violation in rotation.

Stanwix has indicated he may re-evaluate their data in the earth-fixed 
frame, but as of this writing, he has not been able to do so.

\medskip
\noindent
\underline {\textbf{Conclusion}}: In their report, Stanwix et al analyzed 
data in the sidereal frame, so their results are inconclusive for our 
purpose.

\subsection{Herrman et al}
Herrman et al\cite{Herrmann:2005} tested Lorentz invariance by comparing 
resonance frequencies of a continuously rotating optical resonator with a 
stationary one. Although it is not obvious from their report, in personal 
communication, Herrman kindly provided greater detail of their analysis 
procedure, including data plotted in a manner similar to that of Fig. 9. 
Any possible constant lab frame phase signal was significantly less than an 
order of magnitude smaller than that found by Brillet and Hall.

\medskip
\noindent
\underline {\textbf{Conclusion:}} The Herrman et al data, analyzed in the 
earth fixed frame, indicates no anisotropy such as that suggested by the 
anomalous Brillet and Hall signal.

\subsection{Summary of Four Experiments}
As neither the Antonini et al nor Herrmann et al experiments show an 
earth-fixed frame non-null MM type signal, which would suggest local light 
speed anisotropy, it is highly probable that the Brillet and Hall anomalous 
signal truly was attributable to mundane cause. Brillet and Hall did not 
have the advantage of later generation technologies for monitoring such 
things as apparatus tilt, whereas all of the recent researchers did. The 
Stanwix et al experiment is inconclusive as of this writing, though further 
analysis, appropriate to this issue, is expected.

\medskip
\noindent
\underline {\textbf{Conclusion:}} It is all but certain that back and forth 
light speed in rotating frames is isotropic.

\section{Evaluation of Proposed Theories of Rotation}
\label{sec:evaluation}
\subsection{Summary of Pros and Cons}
Table 1 shows the proposed theories of relativistic rotation (see Sec. 
\ref{sec:mylabel3}) vs. their significant characteristics, their 
agreement or lack thereof with key experiments, and their consistency with 
regard to certain theoretical issues.

Note that the Selleri theory considered is his original one, which I believe 
continues to have merit for rotation (though has been subsumed into the 
conventionality thesis for translation.). Question marks signify my 
ignorance with regard to what the particular theory would say, as the topic 
appears to not have been treated in published work by the respective author.

\begin{longtable}[htbp]
{|p{97pt}|p{68pt}|p{60pt}|p{64pt}|p{64pt}|p{64pt}|}
\hline
\endhead
\hline
\endfoot
\underline { }& 
\begin{center}
\underline {\textbf{RRS}} \end{center}  \par \begin{center}
\underline {\textbf{SRT}} \end{center} & 
\begin{center}
\underline {\textbf{Selleri }} \end{center}  \par \begin{center}
\underline {\textbf{Original}} \end{center} & 
\begin{center}
\underline {\textbf{Klauber}} \end{center}  \par \begin{center}
\underline {\textbf{NTO}} \end{center} & 
\begin{center}
\underline {\textbf{Bel}} \end{center}  \par \begin{center}
\underline {\textbf{2 Metrics}} \end{center} & 
\begin{center}
\underline {\textbf{Nikolic}} \end{center}  \par \begin{center}
\underline {\textbf{Local}} \end{center}  \\
\hline
Reference frame& 
\begin{center}
LCIF \end{center} & 
\begin{center}
Local \end{center} & 
\begin{center}
Entire rotating frame \end{center} & 
\begin{center}
Entire rotating frame \end{center} & 
\begin{center}
Local proper frames \end{center}  \\
\hline
Simultaneity& 
\begin{center}
Convention \end{center} & 
\begin{center}
= Lab \end{center} & 
\begin{center}
= Lab \end{center} & 
\begin{center}
? \end{center} & 
\begin{center}
Einstein \end{center}  \\
\hline
Global metric?& 
\begin{center}
No \end{center} & 
\begin{center}
No \end{center} & 
\begin{center}
Yes. \end{center} & 
\begin{center}
Yes. Two: \end{center}  \par \begin{center}
optical {\&} mechanic \end{center} & 
\begin{center}
No. \end{center}  \\
\hline
\begin{center}
\underline {\textbf{Experiment}} \end{center} & 
& 
& 
& 
& 
 \\
\hline
Time dilation?& 
\begin{center}
Yes \end{center} & 
\begin{center}
Yes \end{center} & 
\begin{center}
Yes \end{center} & 
\begin{center}
Yes \end{center} & 
\begin{center}
Yes \end{center}  \\
\hline
Predicts Sagnac? \par (correct $\Delta $ time)& 
\begin{center}
For \textit{$\kappa $} $=v/c$ \end{center}  \par \begin{center}
(or resync) \end{center} & 
\begin{center}
Yes \end{center} & 
\begin{center}
Yes \end{center} & 
\begin{center}
Yes \end{center} & 
\begin{center}
Yes \end{center}  \\
\hline
Antonini/Herrman? (= MM Lorentz contraction)& 
\begin{center}
Yes \end{center} & 
\begin{center}
Yes \end{center} & 
\begin{center}
No \end{center} & 
\begin{center}
No \end{center} & 
\begin{center}
Yes \end{center}  \\
\hline
Brillet {\&} Hall? (= no Lorentz contraction)& 
\begin{center}
No \end{center} & 
\begin{center}
No \end{center} & 
\begin{center}
Yes \end{center} & 
\begin{center}
Yes \end{center} & 
\begin{center}
No \end{center}  \\
\hline
Coriolis effect derivable? (= continuous time)& 
\begin{center}
No \end{center} & 
\begin{center}
Yes \end{center} & 
\begin{center}
Yes \end{center} & 
\begin{center}
? \end{center} & 
\begin{center}
Yes (but discontinuous time) \end{center}  \\
\hline
Byl et al?& 
\begin{center}
Yes \end{center} & 
\begin{center}
Yes \end{center} & 
\begin{center}
Yes \end{center} & 
\begin{center}
? \end{center} & 
\begin{center}
Yes \end{center}  \\
\hline
\begin{center}
\underline {\textbf{Consistent with:}} \end{center} & 
& 
& 
& 
& 
 \\
\hline
Continuous time? \par (clock synched with itself)& 
\begin{center}
Only for \end{center}  \par \begin{center}
\textit{$\kappa $} $=v/c$ \end{center} & 
\begin{center}
Yes \end{center} & 
\begin{center}
Yes \end{center} & 
\begin{center}
? \end{center} & 
\begin{center}
? \end{center}  \\
\hline
Large radius limit case? \par ($\omega r$ const, $\omega \approx $ 0)& 
\begin{center}
Yes, if simultaneity conventional \end{center} & 
\begin{center}
Yes \end{center} & 
\begin{center}
Yes \end{center} & 
\begin{center}
? \end{center} & 
\begin{center}
Yes \end{center}  \\
\hline

\caption{Comparison of Proposed Theories}
\label{tab1}
\end{longtable}

\subsection{Which Theory (Theories) Work(s)?}
\subsubsection{If Brillet and Hall local signal represents reality}
If the anomalous Brillet and Hall signal indicates a genuine attribute of 
physical law, which is doubtful, then only the Bel and Klauber approaches 
would be viable.

\subsubsection{If Antonini/Herrmann data represent reality}
If, as is almost certain, the Antonini et al and Herrmann et al results, 
showing no local anisotropy, represent physical law, then the Bel and 
Klauber approaches, at least in their original incarnations, are not 
correct. As the recent researchers took advantage of innovations in 
monitoring technology not available to Brillet and Hall to measure apparatus 
tilt (the likely cause of the Brillet and Hall non-null local signal), it 
appears prudent to consider the recent experiments as the more 
representative of physical law, in the following consideration of the 
remaining theories.

\medskip
\noindent
\underline {RRS}

The RRS approach embraces conventionality of synchronization in rotation, 
but the analysis of Sec. \ref{subsec:coriolis} shows that only one 
synchronization scheme produces the Coriolis acceleration found in the 
Newtonian limit. This analysis appears to be the death knell for 
conventionality in rotation, and thus also for the RRS approach.

Further, the RRS approach does not provide a global metric, which, it is 
argued in Sec. \ref{sec:mylabel4}, is fundamental for a general 
relativistic description of any physical system.

\medskip
\noindent
\underline {Nikolic}

Nikolic would pass the appropriate Coriolis acceleration test, at least 
locally, and it seems other criteria as well. However, his basic tenet that 
a rotating frame does not represent a single frame, for the purposes of 
general relativistic analysis, appears weak. It does not meet the need, as 
described herein, for a global metric requisite for complete analysis of any 
physical system.

Why, for example, can one not employ the plan of Sec. 
\ref{subsec:mylabel2}? Certainly, one can climb aboard a rotating 
disk, with meter sticks and clocks in hand, lay out a coordinate grid, 
determine a global metric, and use it to analyze any physical situation. I 
consider any system where one can do this as a valid general relativistic 
frame, though Nikolic prefers the local proper frame for studying the 
physical effects viewed by a local observer. While Nikolic's approach may be 
more convenient for less simple spacetimes, for the flat spacetime of 
rotation, I believe the global metric approach to be more convenient, and 
thus preferable.

\medskip
\noindent
\underline {Selleri}

Selleri has the requisite, unique simultaneity necessary to predict the 
correct Coriolis acceleration, and also agrees with Antonini et al (i.e., 
Lorentz contraction exists.) However, his theory, as relates to rotation, 
comprises only a local $x-y-z-t$ frame with a local metric, rather than a global 
frame with global metric.

I suggest that the Selleri approach is valid, but that it lacks a general 
relativistic underlying structure to deal with all problems in rotation 
(i.e., it lacks a global metric.) We shall see that a combination of 
Selleri's approach with Klauber's appears to provide a consistent theory, 
consonant with the preferred experimental result, and suitable for all 
applications.

\medskip
\noindent
\underline {Klauber}

Klauber's approach satisfies all criteria except, importantly, the recent 
experiments. His belief, carried from differential geometry applied to 
purely spatial spaces that the physical component along a spatial axis in 4D 
equals the number of real world meter sticks one would actually use, appears 
to be incorrect. If that one interpretation is modified, Klauber's theory is 
effectively identical to Selleri's, except for the added advantage of being 
applicable, via standard general relativistic methodology, to analysis of 
all problems, kinematic and dynamic, in a rotating frame.

\medskip
\noindent
\underline {\textbf{Conclusion:}} A fully satisfactory theory of rotation 
could comprise a combination of the Selleri and Klauber approaches (see 
Sec. \ref{subsec:mylabel3}). That would agree with the preferred 
MM test result, predict the correct accelerations, have continuous 
single-valued time, and could be used for analysis of any type of problem in 
relativistic rotation. Alternatively, one could use the Klauber approach, 
but modify appropriate results by the Lorentz factor to match real world 
measurement.

\section{Two Suitable Approaches to Relativistic Rotation}
\label{sec:mylabel6}
In this section, we continue to assume that the Antonini et al and Herrmann 
et al results are correct, i.e., circumferential Lorentz contraction exists 
in rotation (See Sec. \ref{subsec:sagnac}.) Thus, in rotation, 
the 4D physical component length along a circumferentially aligned spatial 
axis, e.g., \textit{ds} in the \textbf{e}$_{\phi }$ direction (\textit{ds} in this direction 
equals $d\hat {\phi })$, does not equal the number of actual meter sticks 
one would use in the real world. So, the tried and true differential 
geometry method for finding real world lengths (called physical components) 
in purely spatial spaces is no longer applicable. (See Secs. 
\ref{sec:physical}, \ref{sec:alternative}, 
\ref{subsubsec:space}.)

\subsection{Modified Klauber Approach}
To completely analyze all behavior in rotation, one could use the Klauber 
approach\cite{Klauber:2004}$^{,}$\cite{Klauber:2003}$^{,}$\cite{Klauber:2}$^{,}$\cite{Klauber:2}$^{,}$\cite{Klauber:2001}, 
but simply modify all final results, i.e., all 4D physical component values, 
to accommodate the difference between real world meter stick measurements 
and 4D physical values in the circumferential direction. That is, any 
physical value for any \textit{$\phi $} direction component of a contravariant vector would 
simply be divided by the Lorentz factor $\sqrt {1-\omega ^2r^2/c^2} $. Other 
component real world measured values would simply equal the 4D physical 
value.

For example, the 4D physical length $d\hat {\phi }$ in meters would be 
corrected to equal the real world number of meter sticks, i.e., for
\begin{equation}
\label{eq46}
ds_{circum} =d\hat {\phi }=\sqrt {g_{\phi \,\phi } } d\phi 
=\,\mbox{4D}\,\,\mbox{physical}\,\,\mbox{value},
\end{equation}
Then
\begin{equation}
\label{eq47}
\frac{d\hat {\phi }}{\sqrt {1-\omega ^2r^2/c^2} 
}=\,\mbox{real}\,\,\mbox{world}\,\,\mbox{length}\,\,\mbox{in}\,\,\mbox{meter}\,\,\mbox{sticks}.
\end{equation}
Similarly, for velocity,
\begin{equation}
\label{eq48}
\hat {v}^\phi =\sqrt {g_{\phi \phi } } \,v^\phi , 
\,\,\,\,\,\,\,\,\,v_{real\,world}^\phi =\frac{\hat {v}^\phi }{\sqrt 
{1-\omega ^2r^2/c^2} },
\end{equation}
current density,
\begin{equation}
\label{eq49}
\hat {j}^\phi =\sqrt {g_{\phi \phi } } \,j^\phi ,
\,\,\,\,\,\,\,\,\,j_{real\,world}^\phi =\frac{\hat {j}^\phi }{\sqrt 
{1-\omega ^2r^2/c^2} }
\end{equation}
and any other contravariant vector or tensor component. (For more on which 
type of component is relevant, contravariant or covariant, for a particular 
physical quantity, see ref. \cite{Klauber:2001}.)

For these changes, the one-way velocities calculated by Klauber yield a null 
MM result and the proper Sagnac result. In addition, the correct centrifugal 
and zeroth order Coriolis effects are found as well (with higher order 
corrections to (\ref{eq27}) and (\ref{eq29}).)

The simplicity of this method will become obvious in the next section, where 
we consider an alternative using coordinates whose values reflect the actual 
real world measured values.

\subsection{The Selleri-Klauber Hybrid Approach}
\label{subsec:mylabel3}
\subsubsection{A suitable coordinate choice with Lorentz contraction}
An alternative solution is to selectively choose our normally arbitrary 
coordinate in the circumferential direction \textit{$\phi $} such that \textit{d$\phi $} (rather than \textit{ds}) 
reflects the number of real world meter sticks. More precisely, out of many 
possible choices, we choose a unique coordinate grid for \textit{$\phi $}, such that 
\begin{equation}
\label{eq50}
rd\phi _{SK} 
=\,\mbox{number}\,\mbox{of}\,\mbox{real}\,\mbox{world}\,\mbox{meter}\,\mbox{sticks}\,\mbox{in}\,\mbox{rotating}\,\mbox{frame},
\end{equation}
where we have used the subscript ``SK'', because, as we will see, this 
choice represents an amalgamation of the Selleri and Klauber approaches. The 
form the spatial grid in the rotating frame would take for this choice of 
\textit{$\phi $}$_{SK}$ is illustrated in Fig. 10. Note that, in the lab
\begin{equation}
\label{eq51}
Rd\Phi 
=\,\mbox{number}\,\mbox{of}\,\mbox{real}\,\mbox{world}\,\mbox{meter}\,\mbox{sticks}\,\mbox{in}\,\mbox{lab}.
\end{equation}

\begin{figure}
\centerline{\includegraphics{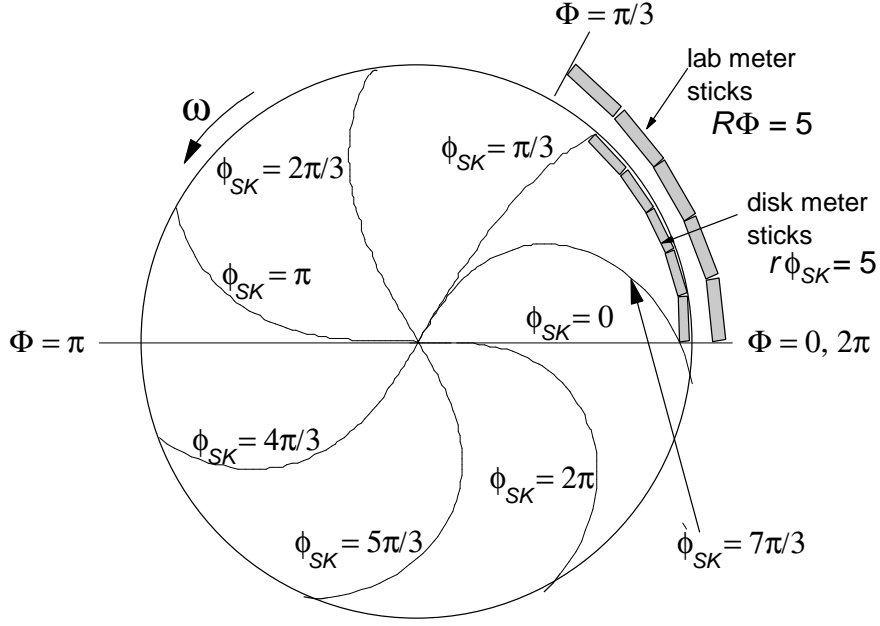}}
\caption{. Selleri-Klauber Hybrid Coordinates}
\label{fig10}
\end{figure}

Further, the 4D physical component in rotation is
\begin{equation}
\label{eq52}
\sqrt {1-\omega ^2r^2/c^2} rd\phi _{SK} 
=\,\mbox{meters}\,\mbox{for}\,\mbox{the}\,\mbox{interval}\,\,ds\,\mbox{in}\,\mbox{rotating}\,\mbox{frame}\,\mbox{coords}.
\end{equation}
This, of course, equals \textit{ds} in the lab between the same two events, which 
equals (\ref{eq51}), i.e.,
\begin{equation}
\label{eq53}
Rd\Phi 
=\,\mbox{meters}\,\mbox{for}\,\mbox{same}\,\mbox{interval}\,\,ds\,\mbox{in}\,\mbox{lab}\,\mbox{coords}.
\end{equation}
\subsubsection{The transformation and the metric}
\label{subsubsec:mylabel2}
Relationship (\ref{eq50}), for a local Selleri type $x-y-z$ spatial grid, equals what 
Selleri calls (with our notation of Secs. \ref{subsec:selleri} 
and \ref{subsubsec:space}) \textit{dx}$_{S}$, and what we have called 
\textit{dx}$_{K}$, the alternative Klauber approach (Sec. 
\ref{subsubsec:space}). Thus,
\begin{equation}
\label{eq54}
rd\phi _{SK} =dx_S =dx_K .
\end{equation}
However, this $x-y-z$ grid approach (middle and right parts of (\ref{eq54})) is limited to 
local frames, and does not yield a global rotating frame metric. To yield a 
fully functional general relativistic approach, we switch to a 
transformation and metric expressed in 1) global cylindrical coordinates, 
and 2) judiciously chosen coordinates, such that $t_{SK}$ and \textit{rd$\phi $}$_{SK}$ equal 
time measured on standard clocks, and length measured with standard meter 
sticks, respectively.

Thus, the transformation from lab (upper case) to rotating frame (lower 
case) coordinates is a hybrid of the Selleri and Klauber transformations, 
i.e.,
\begin{equation}
\label{eq55}
\begin{array}{c}
 ct_{SK} =\sqrt {1-\omega ^2r^2/c^2} cT\quad \quad \quad \quad \quad 
\mbox{(a)} \\ 
 r=R\quad \quad \quad \quad \quad \quad \;\;\;\quad 
\,\,\,\,\,\,\,\,\,\,\,\,\,\,\,\,\,\,\,\,\,\,\,\,\,\,\mbox{(b)} \\ 
 \phi _{SK} =\frac{1}{\sqrt {1-\omega ^2r^2/c^2} }\left( {\Phi -\omega T} 
\right)\quad \;\,\,\,\,\,\,\mbox{(c)} \\ 
 z=Z\quad \quad \quad \quad \quad \quad \quad 
\;\;\;\,\,\,\,\,\,\,\,\,\,\,\,\,\,\,\,\,\,\,\,\,\,\,\,\,\,\,\mbox{(d)}\, \\ 
 \end{array},
\end{equation}
where we have used subscripts only on those coordinates that vary from their 
lab counterparts (and that differ from either the original Selleri or 
Klauber approaches).

The inverse transformation of (\ref{eq55}) is
\begin{equation}
\label{eq56}
\begin{array}{c}
 cT=\frac{ct_{SK} }{\sqrt {1-\omega ^2r^2/c^2} }\quad \quad \quad \quad \quad \quad 
\quad \quad 	\,\,\,\,\,\,\,\mbox{(a)} \\ 
 R=r\quad \quad \quad \quad \quad \quad \quad \;\;\;\quad 
\,\,\,\,\,\,\,\,\,\,\,\,\,\,\,\,\,\,\,\,\,\,\,\,\,\,	\,\,\,\,\,\mbox{(b)} \\ 
 \Phi =\sqrt {1-\omega ^2r^2/c^2} \phi _{SK} +\frac{\omega t_{SK} }{\sqrt 
{1-\omega ^2r^2/c^2} }\quad \;\,\mbox{(c)} \\ 
 Z=z\quad \quad \quad \quad \quad \quad \quad \quad 
\;\;\;\,\,\,\,\,\,\,\,\,\,\,\,\,\,\,\,\,\,\,\,\,\,\,\,\,\,\,	\,\,\,\,\,\mbox{(d)}\,\mbox{.}\, 
\\ 
 \end{array}
\end{equation}
We need to express (\ref{eq56}) in differential form, since by substituting into the 
lab line element, we can find the rotating frame line element, and thus, the 
rotating frame metric. Thus, we find
\begin{equation}
\label{eq57}
\begin{array}{c}
 cdT=\frac{cdt_{SK} }{\sqrt {1-\omega ^2r^2/c^2} }\,\,-\,\,\frac{t_{SK} 
}{\left( {1-\omega ^2r^2/c^2} \right)^{3/2}}\left( {\frac{\omega ^2r}{c^2}} 
\right)dr\quad \quad \quad \quad 	\,\,\,\,\,\,\,\mbox{(a)} \\ 
 dR=dr\quad \quad \quad \quad \quad \quad \quad \;\;\;\quad \quad \quad \quad \quad \quad \quad
\,\,\,\,\,\,\,\,\,\,\,\,\,\,\,\,\,\,\,\,\,\,\,\,\,\,	\,\,\,\,\,			\,\,\,\,\mbox{(b)} 
\\ 
 d\Phi =\sqrt {1-\omega ^2r^2/c^2} d\phi _{SK} -\,\,\frac{\phi _{SK} 
}{\left( {1-\omega ^2r^2/c^2} \right)^{3/2}}\left( {\frac{\omega ^2r}{c^2}} 
\right)dr \\ 
 \,\,\,\,\,\,\,\,\,\,+\frac{\omega dt_{SK} }{\sqrt {1-\omega ^2r^2/c^2} 
}+\frac{\omega t_{SK} }{\left( {1-\omega ^2r^2/c^2} \right)^{3/2}}\left( 
{\frac{\omega ^2r}{c^2}} \right)dr\quad \quad \quad \quad
\;\,	\,\,\,\,\,\,\,\,\,\,\,\mbox{(c)} \\ 
 dZ=dz\quad \quad \quad \quad \quad \quad \quad \quad \quad \quad \quad \quad \quad \quad \quad
\;\;\;\,\,\,\,\,\,\,\,\,\,\,\,\,\,\,\,\,\,\,\,\,\,\,\,\,\,\,	\,\,\,\,\,			\,\,\,\mbox{(d)}\,\mbox{.}\, 
\\ 
 \end{array}
\end{equation}
Substituting, along with $R = r$, into
\begin{equation}
\label{eq58}
(ds)^2=-(cdT)^2+(dR)^2+(Rd\Phi )^2+(dZ)^2,
\end{equation}
one gets a very complicated line element and resulting metric, with the 
latter having many off diagonal terms. From the non-orthogonality between 
lines of constant \textit{$\phi $}$_{SK}$ and $r$ seen in Fig. 10, the \textit{drd$\phi $} terms are to be 
expected. With a little thought, the other terms make sense as well.

We will not carry out the tedium of constructing the metric, but it is 
straightforward, and can be used to solve virtually any problem in rotation. 
Since the coordinate values \textit{dt}$_{SK}$ and \textit{d$\phi $}$_{SK}$ (via \textit{rd$\phi $}$_{SK})$ reflect the 
real world measured values, components of vectors and tensors will directly 
equal the real world values (provided, for example, that one multiplies 
contravariant components in the \textit{$\phi $}$_{SK}$ direction by $r$.)

We will do something a little simpler, computation-wise, by considering \textit{dr = dz} = 
0, i.e., motion along a circumferential line. For this case, (\ref{eq58}) and (\ref{eq57}) 
yield the line element
\begin{equation}
\label{eq59}
(ds)^2=-(cdt_{SK} )^2+r^2\left( {1-\omega ^2r^2/c^2} \right)(d\phi _{SK} 
)^2+2r^2\omega d\phi _{SK} dt_{SK} .
\end{equation}
Considering light with \textit{ds = }0, and using the quadratic equation formula to solve 
(\ref{eq59}) for the one-way speed of light \textit{rd$\phi $}$_{SK}$/\textit{dt}$_{SK}$, we find
\begin{equation}
\label{eq60}
v_{real\,world,\mbox{ }circum,\mbox{ }1-way} =\frac{rd\phi _{SK} }{dt_{SK} 
}=\frac{\omega r\pm c}{1-\omega ^2r^2/c^2},
\end{equation}
where the sign in the numerator before $c$ indicates the direction of the light 
ray, ccw or cw. 

Using (\ref{eq60}), one can find the back and forth, MM time to be
\begin{equation}
\label{eq61}
t_{MM} =\frac{l}{\left| {v_+ } \right|}+\frac{l}{\left| {v_- } 
\right|}=\frac{2l}{c},
\end{equation}
and thus, the back and forth speed of light along the circumference is $c$.

Similarly the Sagnac time difference for cw and ccw rays, with $l=2\pi 
r/\sqrt {1-\omega ^2r^2/c^2} $, is
\begin{equation}
\label{eq62}
t_{Sagnac,\,real\,world} =\frac{l}{\left| {v_- } \right|}-\frac{l}{\left| 
{v_+ } \right|}=\frac{4\pi \omega r^2}{c^2\sqrt {1-\omega ^2r^2/c^2} },
\end{equation}
in agreement with (\ref{eq4}).

For other, more general problems, due to the complicated nature of the 
metric found from (\ref{eq57}) and (\ref{eq58}), the modified Klauber approach appears far 
simpler computationally. As but one example, the Coriolis calculation for 
both approaches will be the same, but the modified Klauber method is far 
easier.

\section{Loose Ends}
\label{sec:loose}
\subsection{Curvature: Meter Stick Measurement or Spacetime Path?}
The conclusions of the prior section lead to an interesting result with 
regard to curvature of a rotating disk surface. Using meter stick 
measurements of the circumference, one would then find
\begin{equation}
\label{eq63}
\mbox{circumference}\,\,>2\pi r\,\,\,\,\,\left( 
{\mbox{measured}\,\mbox{with}\,\mbox{disk}\,\mbox{meter}\,\mbox{sticks = 
}\Delta x_s } \right)
\end{equation}
and thus would be inclined to deem the surface curved.

However, curvature in differential geometry is defined via the variation in 
the interval \textit{ds}, not the coordinate value \textit{dx}$_{S}$. For this, and continuous, 
single-valued time (such as that of the Selleri and Klauber approaches, 
which lead to correct Coriolis acceleration), we would have
\begin{equation}
\label{eq64}
\mbox{circumference}\,\,=2\pi r\,\,\,\,\,\left( 
{\mbox{spacetime}\,\,\mbox{interval}\,\,\Delta s} \right).
\end{equation}
The curvature found from the Riemann tensor is based on \textit{ds} variation, \textit{not} 
\textit{dx}$_{S}$ variation. Thus, Riemann curvature for the 2D disk surface is zero.

Another way to the same conclusion is to recognize that (\ref{eq56}) with $t_{SK}$ = 
$z$ = 0 is simply a transformation from the 2D lab surface to the 2D rotating 
surface, and since the Riemann tensor for this surface is zero in the lab, 
it must also be zero for the rotating surface.

\medskip
\noindent
\underline {\textbf{Conclusion:}} For continuous, single-valued time (i.e., 
correct Coriolis acceleration), the Riemann curvature for the rotating disk 
surface is zero, even if real world, rotating frame meter stick measurement 
of the circumference does not equal 2$\pi r$.

\subsection{Resolution of the Limit Paradox}
\label{subsec:resolution}
The Selleri limit paradox was re-cast in Sec. 
\ref{subsubsec:selleri} to the following.

Given that (in order to predict correct Coriolis acceleration) time in the 
rotating frame must be continuous, then only a single simultaneity is 
allowed in that frame. Yet at Selleri's limit, the LCIF, which is a 
translating frame, can have any of an infinite number of allowable 
simultaneity schemes, and thus is not equivalent locally to the rotating 
frame at that location.

As discussed in Ref. \cite{Klauber:2}, the potential -- 
$\raise.5ex\hbox{$\scriptstyle 1$}\kern-.1em/ 
\kern-.15em\lower.25ex\hbox{$\scriptstyle 2$} \omega ^{2}r^{2}$ does 
not go to zero in the Selleri limit, and relativistic mass depends on total 
energy, kinetic plus potential. Hence, in the rotating frame at the Selleri 
limit, mass of a proton, for example, would be different than the mass of a 
proton in the LCIF. Thus, by simply measuring the mass of Avogadro's number 
of carbon atoms, for example, in each frame, one could discern which of the 
two were rotating. Thus, they are not equivalent.

\medskip
\noindent
\underline {\textbf{Conclusion:}} Change in mass, due to potential energy, 
in a rotating frame permits one to distinguish between that frame and an 
LCIF frame.

\section{Summary of Conclusions}
\label{sec:summary}
The most important result of this article is that correct Newtonian limit 
Coriolis acceleration can only be predicted by general relativity for a 
unique simultaneity in the rotating frame, where this simultaneity equals 
that of the lab and ensures continuity of time (single valued time, with 
clocks in synchronization with themselves.) I submit this settles what has 
probably been the most contentious issue in relativistic rotation. We list 
this conclusion with others below.

\subsection{Lorentz Contraction}
\begin{itemize}
\item The widely dispersed logic justifying circumferential Lorentz contraction in a rotating frame is submitted to contain unresolved inconsistencies.
\item Contraction, by the standard definition for which end points of a meter stick are simultaneous, varies with choice of synchronization scheme, so Lorentz contraction (by the specific Lorentz factor \textit{$\gamma $}) and conventionality of synchronization in rotation cannot both be true.
\item Proof for Lorentz contraction in rotation was found in both the Antonini et al and Herrmann et al experiments, which conflicted with an earlier anomalous signal found by Brillet and Hall.
\item The rotating disk surface has a circumference, as measured by real world meter sticks greater than 2$\pi r$, but an interval $\Delta s$ equal to 2$\pi r$, and thus its 2D Riemann curvature is zero.
\end{itemize}
\subsection{Simultaneity/Synchronization}
\begin{itemize}
\item Coriolis acceleration arises in the general relativistic equation of motion from the off diagonal time-space term in the metric. This term is dependent upon choice of simultaneity.
\item Correct Newtonian limit Coriolis acceleration only arises from one simultaneity choice, the one for which time is continuous (and clocks are in synchronization with themselves), and thus, there is no conventionality of synchronization in rotation.
\end{itemize}
\subsection{Locality and the Selleri Limit ``Paradox''}
\begin{itemize}
\item Due to relativistic mass changes from potential energy, the rotating frame is not locally equivalent to a LCIF.
\item The newly formulated version of the Selleri limit ``paradox'' is thus resolved, as the LCIF and rotating frames in the limit are not equivalent.
\end{itemize}
\subsection{Comparison of Theories}
\begin{itemize}
\item The traditional LCIF approach to rotation, in its most modern version, the RRS approach, embraces conventionality of synchronization and thus, for reasons delineated herein, is argued to be incorrect. Further, it provides no global metric and therefore is incapable of being a generally applicable approach to all problems in rotation, such as, for one example, Coriolis acceleration.
\item The original Selleri approach has the correct simultaneity and agrees with the Antonini et al, Herrmann et al, and Sagnac experiments. However, it does not provide a global metric, and thus is also not fully applicable to general problems.
\item Bel's approach predicts a non-null MM result and thus disagrees with the recent experiments. Further, his theory appears to be in conflict with the Byl et al experiment.
\item Nikolic predicts the correct Newtonian Coriolis acceleration and a null MM signal, but his global (i.e. integrated) time appears discontinuous, and he provides no global metric. Further, his argument that there is no single rotating frame, but only an infinite series of local frames, appears difficult to justify.
\item Klauber has continuous time and the correct Coriolis acceleration, but disagrees with the Antonini et al and Herrmann et al experiments. His prediction of no rotational Lorentz contraction (based on an interpretation of the standard physical component analysis method of differential geometry) is almost certainly incorrect.
\item None of the proposed theories of rotation makes correct predictions for all cases.
\end{itemize}
\subsection{Two Suitable Theories of Rotation}
\begin{itemize}
\item A modified Klauber theory, for which contravariant physical components in the circumferential direction are modified by the Lorentz factor is simple, generally applicable to rotation, and provides the correct results for all cases.
\item A hybrid Selleri-Klauber theory, in which Selleri's ``built-in'' Lorentz contraction is incorporated by selecting one particular spatial coordinate grid, provides the correct simultaneity, as well as a global metric with the capability of analyzing all problems.
\end{itemize}
\section{Acknowledgements}
I thank two anonymous reviewers for suggesting changes that markedly 
clarified the presentations of several sections.

\end{document}